\documentclass[%
 reprint,
 amsmath,amssymb,
 aps,
]{revtex4-2}

\usepackage{graphicx}% Include figure files
\usepackage{dcolumn}% Align table columns on decimal point
\usepackage{bm}% bold math
\usepackage{xcolor}
\usepackage{hyperref}% add hypertext capabilities
\hypersetup{
  colorlinks   = true, 
  urlcolor     = blue, 
  linkcolor    = blue, 
  citecolor   = blue 
}
\usepackage{bbm}
\usepackage{verbatim}
\usepackage{abraces} % advanced brace handling
\usepackage[normalem]{ulem}

\newcommand{\tr}[1]{\operatorname{tr}\left[#1\right]}
\newcommand{\abs}[1]{\lvert #1 \rvert}
\newcommand{\rket}[1]{\lvert #1 )}
\newcommand{\rbra}[1]{( #1 \rvert}
\newcommand{\ket}[1]{\lvert #1 \rangle}
\newcommand{\bra}[1]{\langle #1 \rvert}
\newcommand{\braket}[2]{\langle #1 \vert #2 \rangle}
\newcommand{\hmax}{h_{\mathrm{max}}}
\newcommand{\hypf}{\,_2F_1}

\begin{document}

%\preprint{APS/123-QED}

\title{From Dual Unitarity to Generic Quantum Operator Spreading}% Force line breaks with \\

\author{Michael A. Rampp, Roderich Moessner, and Pieter W. Claeys}
\affiliation{Max Planck Institute for the Physics of Complex Systems, 01187 Dresden, Germany}

\date{\today}% It is always \today, today,

\begin{abstract}
Dual-unitary circuits are paradigmatic examples of exactly solvable yet chaotic quantum many-body systems, but solvability naturally goes along with a degree of non-generic behaviour. By investigating the effect of weakly broken dual unitarity on the spreading of local operators we study whether, and how, small deviations from dual unitarity recover fully generic many-body dynamics.
We present a discrete path-integral formula for the out-of-time-order correlator and recover a butterfly velocity smaller than the light-cone velocity, $v_B < v_{LC}$, and a diffusively broadening operator front, two generic features of ergodic quantum spin chains absent in dual-unitary circuit dynamics. 
The butterfly velocity and diffusion constant are determined by a small set of microscopic quantities and the operator entanglement of the gates has a crucial role. 
\end{abstract}

\maketitle

%\tableofcontents

The dynamical behaviour of strongly correlated quantum many-body systems out of equilibrium is notoriously hard to describe. 
Quantum many-body dynamics is intimately related to questions of thermalization, information scrambling, quantum chaos, and the emergence of hydrodynamics \cite{Rigol2008,Khemani2018,Chan2018,khemani2018velocity,von2018operator,nahum2018operator,xu2019locality,xu2020accessing}. 
Simple model systems that can be solved analytically are highly desirable since they offer an invaluable window into these questions. In recent years dual-unitary circuits (DUCs) have emerged as paradigmatic examples of exactly solvable yet chaotic many-body systems~\cite{akila2016particle,bertini2019exact,claeys2021ergodic,aravinda2021dual,Gopalakrishnan2019,bertini2019entanglement,bertini2020operator,bertini2018exact,claeys2020maximum,Bertini2020,piroli2020exact,lerose2021influence,Jonay2021,Suzuki2022,Claeys2022a,Borsi2022,Kasim2023,Kos2022,Stephen2022} in which a variety of dynamical quantities are analytically accessible~\cite{bertini2019exact,claeys2021ergodic,aravinda2021dual,Gopalakrishnan2019,bertini2019entanglement,bertini2020operator,claeys2020maximum,Bertini2020}. 
However, solvability comes at a cost in genericity, and in several respects DUCs display behaviour that differs strikingly from the phenomenology observed numerically in more generic models~\cite{von2018operator,nahum2018operator,khemani2018velocity,xu2019locality,Zhou2020,xu2020accessing,LopezPiqueres2021}. In particular, the space-time duality present in DUCs enforces that correlations and operators spread with the maximum possible velocity. This nongeneric maximal spreading has also been observed experimentally on Google's quantum processor~\cite{mi2021information}.

It is natural to ask whether and in what sense DUCs might serve as a starting point to understand the behavior of more general systems. This question is especially relevant given recent advances in noisy intermediate scale quantum devices. 
While much is now known about DUCs themselves, there are only few results about deviations from dual unitarity~\cite{kos2021correlations,zhou2022maximal,Ippoliti2022} and many questions remain open. In Ref.~\cite{kos2021correlations} the behavior of local correlation functions in circuits close to dual unitarity has been investigated. 
It was found that in most instances local correlators acquire a more generic spatio-temporal structure, not being exclusively supported on the light cone anymore.

In this paper we investigate operator dynamics by considering out-of-time-order correlators (OTOCs) in a broad class of chaotic quantum circuits in which dual unitarity is weakly broken. 
We show that the OTOC can be expressed as a sum over all possible paths resulting from scattering on individual dual-unitarity-breaking gates, acting as defects in an otherwise dual-unitary circuit.
We find that after an initial period in which the dual-unitary form is approximately preserved, even a weakly broken duality leads the OTOC to recover a nonmaximal butterfly velocity and a diffusive broadening of the operator front, hallmarks of generic one-dimensional chaotic quantum many-body systems absent from pure DUCs.
The operator front at late times takes a universal form and its parameters are microscopically determined by the entangling properties of the gate. Our manipulations are controlled in the  limit where the dual-unitarity-breaking gates are dilute along the direction of the light cone, but we argue that they capture the relevant characteristics of the OTOC on intermediate to long timescales even in the case of a Floquet circuit with dense perturbations. The developed framework is expected to be applicable to different probes of operator spreading.

\begin{figure}[tb]
    \centering
    \includegraphics[width = 0.45\textwidth]{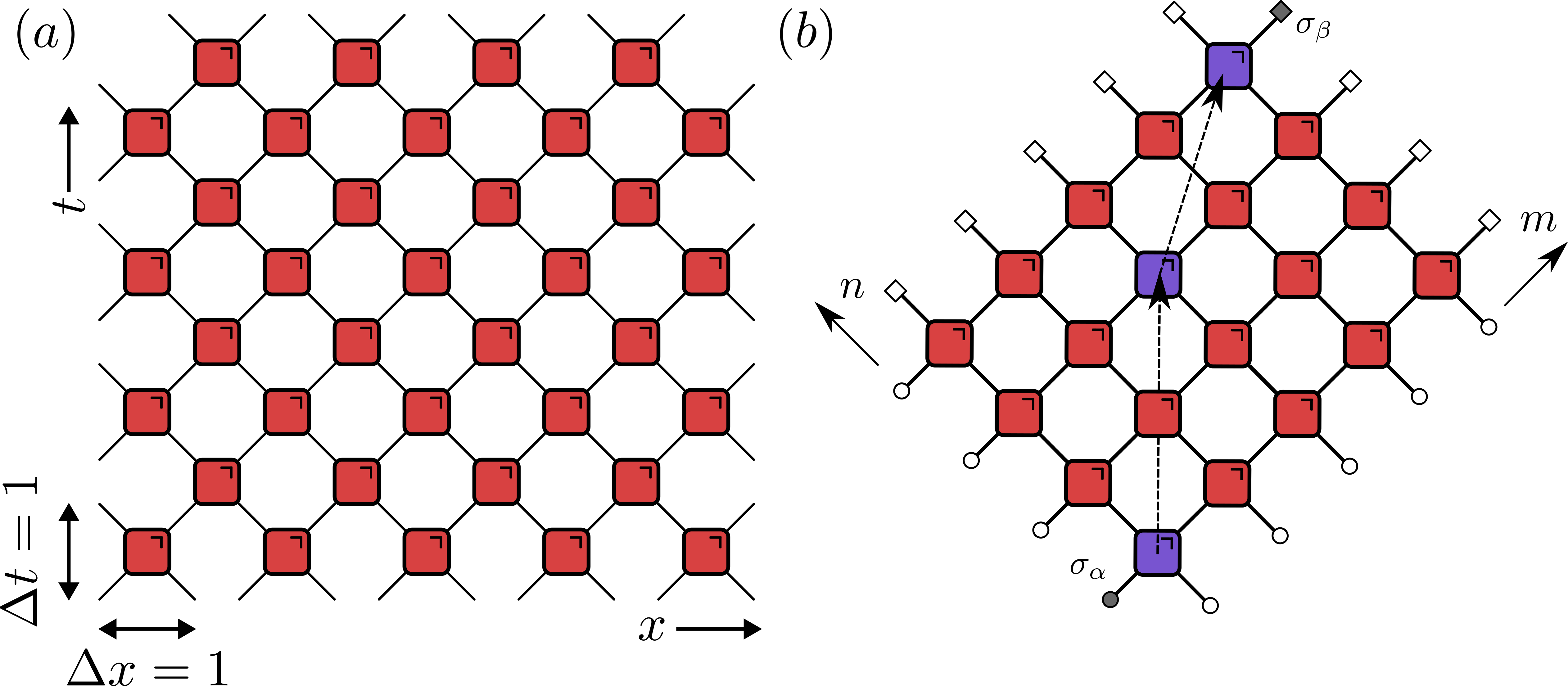}
    \caption{(a) The unitary time evolution operator is constructed as a brickwork circuit composed out of identical two-site gates. One row of gates corresponds to a time step $\Delta t=1$. (b) Graphical depiction of processes contributing to the OTOC. When dual unitarity is broken, the edge of the operator string can scatter into the light cone. The OTOC is then given by a weighted sum of all paths. 'Folded' dual-unitary gates are depicted in red and dual-unitary-breaking gates in purple.}
    \label{fig:circuits}
    \vspace{-1\baselineskip}
\end{figure}

\emph{Dual-unitary circuits.} We consider circuits composed of unitary gates $U$ acting on two sites with local Hilbert space dimension $q$, with matrix elements $U_{ab,cd}$ graphically expressed as 
\begin{align}
\vcenter{\hbox{\includegraphics[width=0.8\columnwidth]{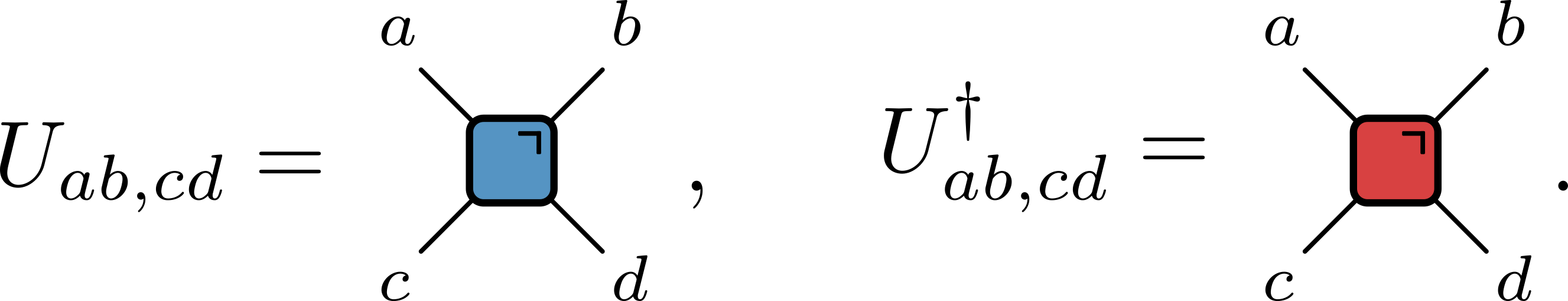}}} \label{eq:def_U}
\end{align}
In this notation each leg corresponds to an index in the local Hilbert space, and connecting two indices corresponds to a tensor contraction (see e.g. \cite{Orus2014}). 
Unitary gates arranged in a brickwork geometry [Fig.~\ref{fig:circuits}(a)] provide simple models for local, unitary quantum many-body dynamics on a one-dimensional lattice \cite{Chan2018,khemani2018velocity,von2018operator,nahum2018operator,Fisher2022}, with the number of discrete time steps $t$ corresponding to the number of rows in the circuit.
A gate $U$ is called \emph{dual-unitary} if the associated dual %(or reshaped) 
gate $\tilde{U}$ defined by $\tilde{U}_{ab,cd}\equiv U_{db,ca}$ is also unitary \cite{Gopalakrishnan2019,bertini2019exact}. %A brickwork circuit composed of dual-unitary gates is unitary not only in the time but also in the space direction.

Let us review the computation of OTOCs in dual-unitary circuits~\cite{claeys2020maximum}. We consider a basis of local operators $\sigma_\alpha$ normalized according to $\tr{\sigma_\alpha^{\dagger}\sigma_\beta}=q\delta_{\alpha\beta}$. Setting $\sigma_0=1$, the remaining operators are traceless, and we take $\sigma_{\alpha}(x)$ to act as $\sigma_\alpha$ on site $x$ and as the identity everywhere else. Denoting the time evolution operator as $\mathcal{U}(t)$, we write $\sigma_{\alpha}(x,t) = \mathcal{U}(t)^\dagger\sigma_\alpha(x)\,\mathcal{U}(t)$
and consider the OTOC
\begin{align}
C_{\alpha\beta}(x,t) =  \langle \sigma_{\alpha}(0,t)\sigma_\beta(x,0)\sigma_{\alpha}(0,t)\sigma_\beta(x,0)\rangle, \label{eq:otoc}
\end{align}
with respect to the maximally mixed state, $\langle O\rangle\equiv\operatorname{tr}[O]/\operatorname{tr}[\mathbbm{1}]$. This function quantifies the spreading of operators and the scrambling of information into nonlocal degrees of freedom~\cite{larkin1969quasiclassical,qi2019quantum,parker2019universal}, and is experimentally accessible in (digital) quantum simulation platforms~\cite{zhu2016measurement,swingle2016measuring,garttner2017measuring,vermersch2019probing,mi2021information,zhao_probing_2022}.
The OTOC's propagation speed is called \emph{butterfly velocity} $v_B$ and sets the maximal speed at which information can spread in the circuit~\cite{bravyi2006lieb,hosur2016chaos}.

The OTOC exhibits a strong parity dependence. Here, we focus on $C^+(x,t)$ for which $(t-x)\in2\mathbb{Z}$, but the derivation is analogous for $C^-(x,t)$, for which $(t-x)\in2\mathbb{Z}+1$. In general circuits, Eq.~\eqref{eq:otoc} can be graphically represented as the contraction of a two-dimensional tensor network, the size of which is set by the light-cone coordinates $n=(t-x+2)/2,\,m=(t+x)/2$,
\begin{align}
C^+_{\alpha\beta}(n,m) &= \frac{1}{q^{n+m}}\,\vcenter{\hbox{\includegraphics[width=0.55\columnwidth]{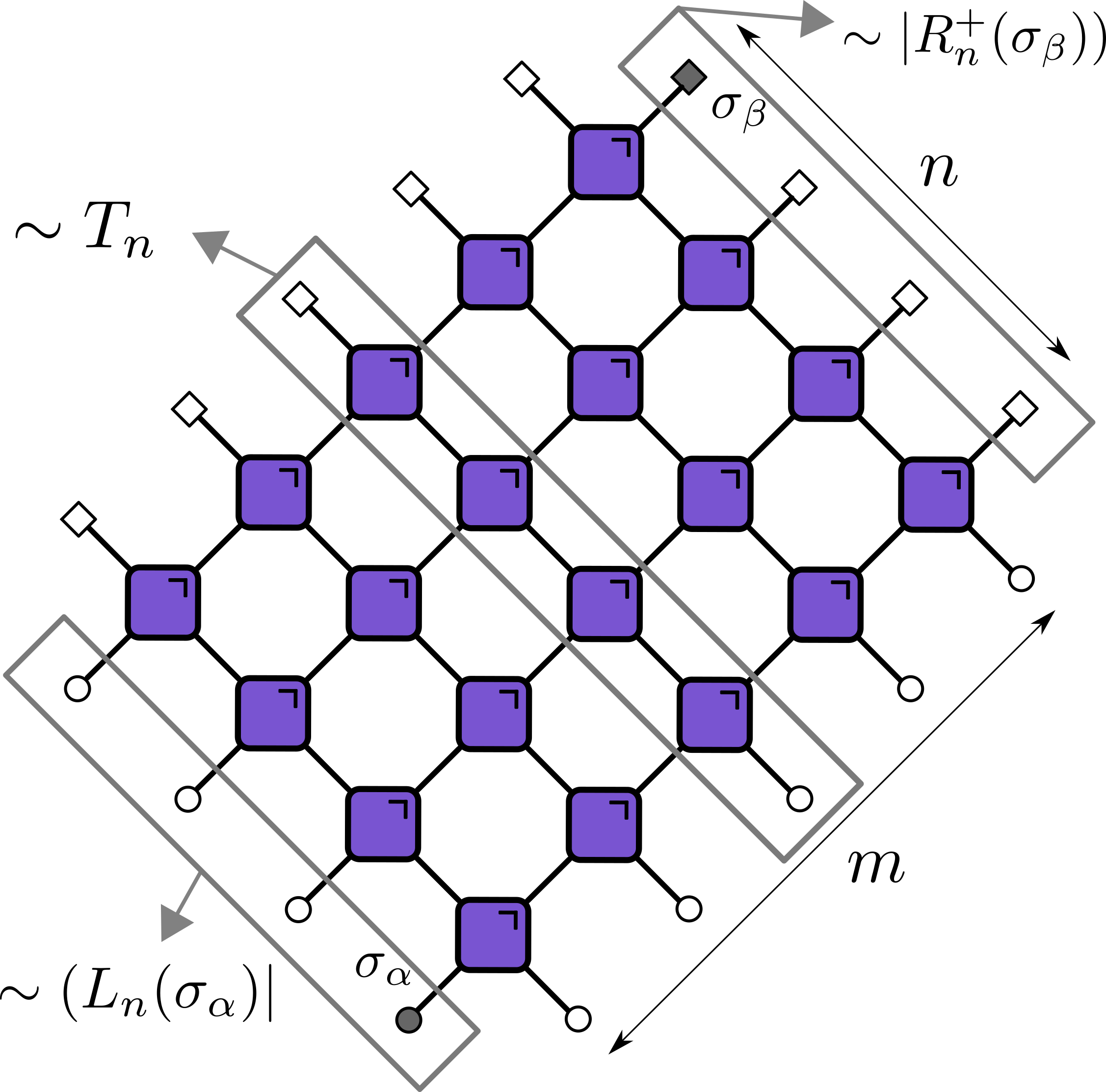}}}\nonumber ,\\
&= (L_n(\sigma_\alpha)\vert (T_n)^m \vert R_n^+(\sigma_\beta)),\label{fig:otoc_plus}
\end{align}
where we have introduced the `folded' gate acting on four copies of the local Hilbert space,
\begin{align}
\vcenter{\hbox{\includegraphics[width=0.1\columnwidth]{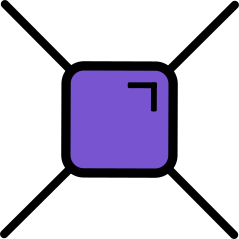}}}\,= \Bigg(\,\, \vcenter{\hbox{\includegraphics[width=0.1\columnwidth]{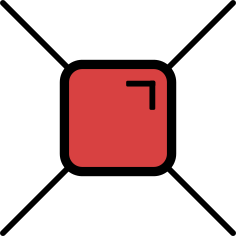}}}  \otimes \vcenter{\hbox{\includegraphics[width=0.11\columnwidth]{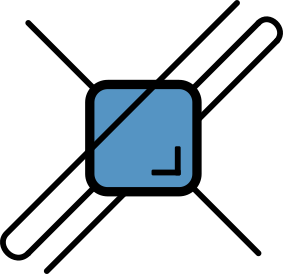}}}& \,\Bigg)^{\otimes2} \label{fig:folded_gate}\,, 
\end{align}
as well as the following vectors,
\label{fig:folded_vectors}
\begin{align}
\vcenter{\hbox{\includegraphics[height=0.025\textheight]{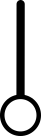}}} \,= \frac{1}{q}\, \vcenter{\hbox{\includegraphics[height=0.025\textheight]{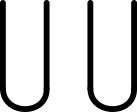}}}\,, \,\, \vcenter{\hbox{\includegraphics[height=0.025\textheight]{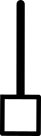}}}\, = \frac{1}{q}\,\vcenter{\hbox{\includegraphics[height=0.025\textheight]{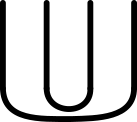}}}\,, \,\,
\vcenter{\hbox{\includegraphics[height=0.032\textheight]{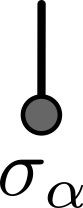}}} \,= \frac{1}{q}\, \vcenter{\hbox{\includegraphics[height=0.026\textheight]{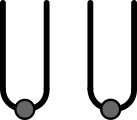}}}\,, \,\, \vcenter{\hbox{\includegraphics[height=0.032\textheight]{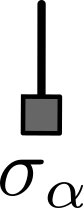}}} \,= \frac{1}{q}\, \vcenter{\hbox{\includegraphics[height=0.032\textheight]{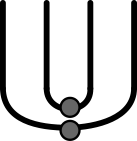}}}\,.
\end{align}
The tensor network \eqref{fig:otoc_plus} can be understood as the contraction of powers of a transfer matrix 
\begin{align}
    T_n = \frac{1}{q}\, \underbrace{\vcenter{\hbox{\includegraphics[height=0.05\textheight,angle=180]{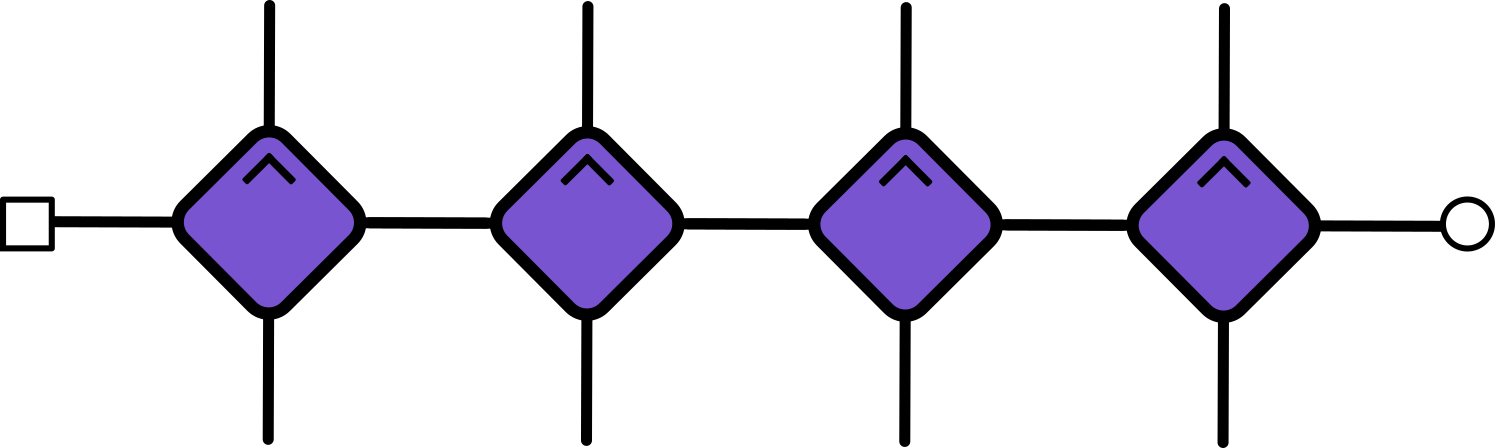}}}}_n\,.
\end{align}
This transfer matrix is a contracting map, i.e. $\| T_n v\|\leq\|v\|$. Hence, all its eigenvalues lie inside or on the boundary of the complex unit disk. In the limit  $m\rightarrow\infty$ the OTOC is completely determined by the eigenvectors of $T_n$ with leading eigenvalue (i.e., modulus 1).
If the gate is dual-unitary there exist $n+1$ leading eigenvectors 
that are independent of any further characteristics of the gate and can be constructed explicitly~\cite{claeys2020maximum}. If these vectors exhaust the set of leading eigenvectors, the gate is called \emph{maximally chaotic}~\cite{bertini2020operator}.

While the existence of a local conserved quantity implies the existence of further leading eigenvectors, the set of maximally chaotic gates is dense in the set of dual-unitary gates~\cite{bertini2020operator}. We call the subspace spanned by this generic set of vectors the \emph{maximally chaotic subspace} (MCS). Its construction and properties are elaborated in the supplemental material~\cite{SuppMat}. 

For circuits composed of maximally chaotic dual-unitary gates the following holds~\cite{claeys2020maximum}:
After an initial transient regime, the OTOC for $(t-x)\in2\mathbb{Z}$ 
is only nonvanishing on the light cone edge, $|x|=t$, where it takes the universal value $-1/(q^2-1)$, resulting in a maximal butterfly velocity $v_B=v_{LC}=1$. For $(t-x)\in2\mathbb{Z}+1$ the OTOC decays exponentially inside the light cone.
This behavior is to be contrasted with the OTOC in generic unitary dynamics, where $v_B < 1$ and the ballistic spreading is accompanied by a diffusively broadening front~\cite{von2018operator,nahum2018operator}.

\emph{Breaking dual unitarity.} 
For concreteness we consider gates of the form $U=Ve^{i\varepsilon W}$ 
where $V$ is a maximally chaotic dual-unitary gate, $W$ is Hermitian, and $\varepsilon$ is taken to be small $(\varepsilon\ll 1)$.

How does the breaking of dual unitarity affect the OTOC? In the absence of dual unitarity and without further constraints, the transfer matrix has only a single, `trivial', leading eigenvector that leads to $\lim_{m\rightarrow\infty}C^+_{\alpha \beta}(n,m)=1$ for finite $n$~\cite{claeys2020maximum}. Signaling that the butterfly velocity in such a circuit is nonmaximal, $v_B<1$, the butterfly velocity then needs to be determined from the subleading eigenvectors of the transfer matrix. 

The contribution of a subleading eigenvector with eigenvalue $\abs{\lambda}<1$ decays on a timescale $\tau \sim\left(-\log\abs{\lambda}\right)^{-1}$. For weak perturbations from dual unitarity, the manifold of leading eigenvectors is split by an amount $\mathcal{O}(\varepsilon^2)$. If in the limit $m \to \infty$ a finite spectral gap to the remaining spectrum exists, then for sufficiently small $\varepsilon$ the largest subleading eigenvectors are predominantly composed out of vectors in the MCS, and the behavior of the OTOC on long times $t\geq 1/\varepsilon^2$ is  determined by those eigenvectors. 
In the absence of symmetries that force a gap closing, we expect this to be the generic scenario.

To proceed, we project the transfer matrix to the MCS, 
similar in spirit to degenerate perturbation theory,  and compute the OTOC with the projected transfer matrix. If the perturbed gates are dilute along the light-cone, this approximation is controlled~\cite{SuppMat}. 
However, 
for perturbations that are sufficiently small compared to the spectral gap we expect this description to remain valid in the dense limit. %, when each gate breaks dual unitarity. 

Notably, the MCS only grows linearly with the size of the transfer matrix compared to the full operator space, which grows exponentially. Reducing the dynamics to the MCS presents a significant computational advantage. Moreover, the resulting transfer matrix can be efficiently truncated, allowing for analytic evaluation (see below).

The matrix elements of the transfer matrix in the MCS, 
$(\ell\vert T_n\vert k)$, can be readily calculated~\cite{SuppMat}. These depend on the properties of the gates through a set of quantities $B_k$, $k=1,\dots,n$.  
Graphically
\begin{align}
B_k = \frac{1}{q^{k+1}}\,\,&\vcenter{\hbox{\includegraphics[width=0.4\columnwidth]{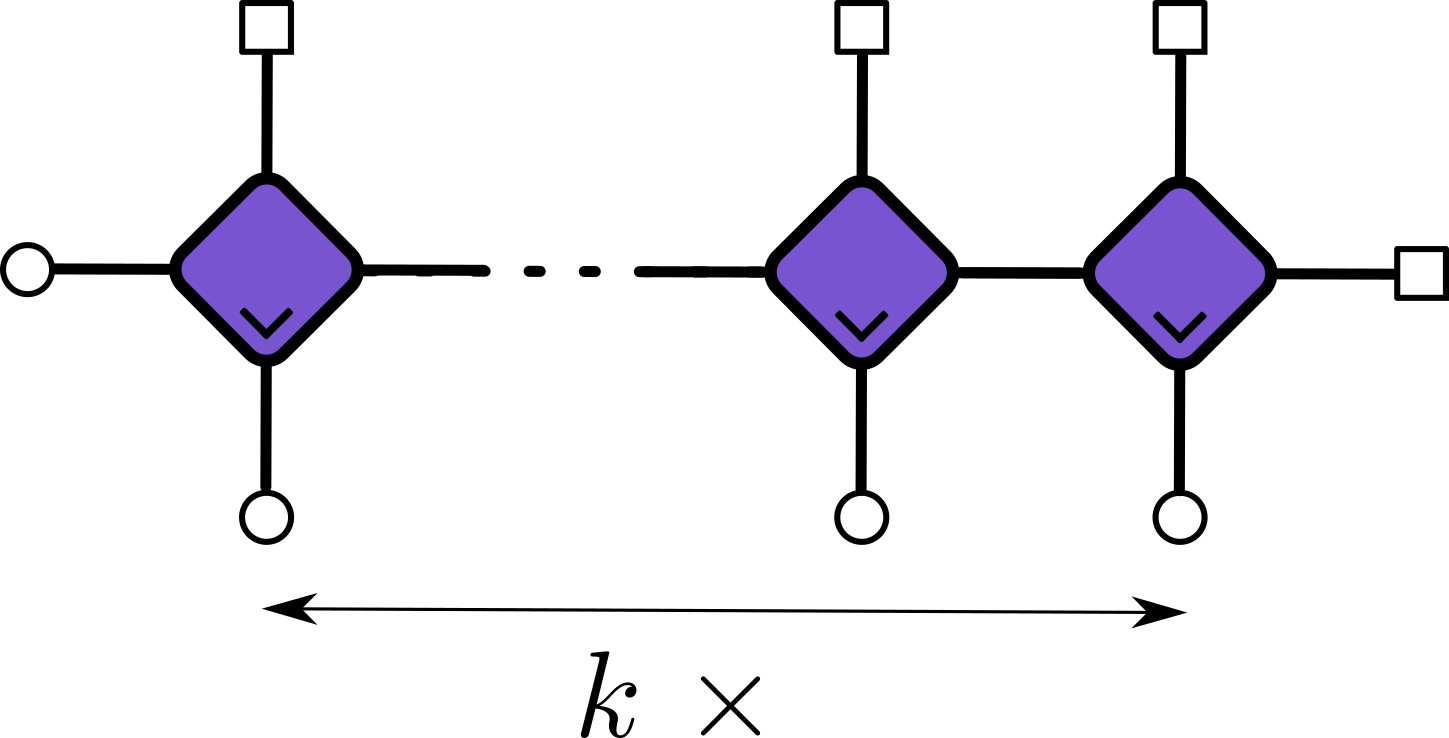}}} \label{fig:gen_bubble}\,.
\end{align}
Defining $z_1:=(B_1-1)/(q^2-1)$ and $z_k:=(B_k-B_{k-1})/(q^2-1)$ for $k>1$  we find that $(0\vert T_n\vert 0) = 1$ and
\begin{subequations}
\label{eq:matrixelements}
    \begin{align}
        (0\vert T_n\vert k) &= \sqrt{q^2 -1}\frac{z_k}{q^{k-1}}, \quad k\geq1 \\
        (\ell\vert T_n\vert k) &= \begin{cases} \frac{q^2 z_{k-l} - z_{k-l+1}}{q^{k-l}}, &\,\, k>l, \\ 1-z_1, &\,\, k=l, \\ 0, &\,\, \mathrm{otherwise}. \end{cases}\label{eq:matrix2}
    \end{align}
\end{subequations}
They have the following structure: (i) the matrix is upper triangular, as a direct consequence of unitarity~\cite{SuppMat}.  (ii) Except in the first row, the $k$-th side diagonal has the same entry everywhere. This is only the case if all gates are identical and follows from translational invariance. Taken together, these  imply that the eigenvalues are $\lambda_0=1$ with algebraic multiplicity $1$, and $\lambda_{1}=1-z_1<1$ with algebraic multiplicity $n$. 

All matrix elements can be given a quantum-information theoretic interpretation. Inserting the Schmidt decomposition of the gate, $U=\sum_j \sqrt{\sigma_j}X_j\otimes Y_j$, into $z_1$ 
reveals that this quantity is equivalent to the linear operator entanglement of the gate $E(U)$~\cite{zanardi2001entanglement}, with $z_1=1-\frac{q^2}{q^2-1}E(U)$. The properties of the operator entanglement imply $0\leq z_1\leq 1$.  As dual unitarity of $U$ is equivalent to $U$ having maximal operator entanglement~\cite{rather2020creating}, $z_1=0$ iff $U$ is dual unitary, and $z_1$ hence quantifies proximity to dual unitarity. 

The subleading eigenvalues follow from Eq.~\eqref{eq:matrix2} as $1-z_1$, such that the timescale $\tau$ for deviations from dual unitarity to become apparent follows as
\begin{equation}\label{eq:timescale}
    \tau^{-1} \sim -\log\left(1-z_1\right)=-\log\left(\frac{q^2}{q^2-1}E(U)\right). 
\end{equation}

All $B_k$ for $k>1$ can analogously be given the interpretation as operator entanglements of $k$-fold diagonally composed gates on 
enlarged Hilbert spaces~\cite{SuppMat}. Because the diagonal composition preserves dual unitarity~\cite{Borsi2022}, $z_k=0$ for all $k$ iff $U$ is dual-unitary. 
The $z_k$ are bounded by $0\leq z_k\leq(1-z_1)^{k-1}$, implying that 
$z_k\rightarrow0$ for $k\rightarrow\infty$~\cite{SuppMat}. 
In general, knowledge of all higher-order operator entanglements is necessary to compute the OTOC, however they are increasingly less important.

While it is not possible to calculate arbitrary powers of the transfer matrix exactly, the structure of Eq.~\eqref{eq:matrixelements} allows for a systematic expansion of the OTOC using a path-integral approach. We write,
\begin{align}
    T_n &= D + \sum_{j=1}^n \rket{u_j}\rbra{j}, \\ D&=\operatorname{diag}\left(1,1-z_1,\dots,1-z_1 \right), \label{eq:tm_decomposition}
\end{align}
where the $\rket{u_j}$ are vectors containing the off-diagonal matrix elements, $(\ell\vert u_k)=(\ell\vert T_n\vert k)$ for $k>\ell$ and $(\ell\vert u_k)=0$ otherwise. 

The off-diagonal terms are of order $z_k\sim\epsilon^2$,  and we expand the OTOC in powers of $(T_n-D)$. Each such power can be expressed as
\begin{align}
    &(T_n - D)^{\nu} = \sum_{j_1, j_2, \dots, j_{\nu}=1}^n (\rket{u_{j_1}} \rbra{j_1}) (\rket{u_{j_2}} \rbra{j_2})\dots  (\rket{u_{j_\nu}} \rbra{j_\nu}) \nonumber \\
    &\quad=\sum_{j_1 < j_2 \dots < j_{\nu}}^n \rket{u_{j_1}} (j_1 \vert u_{j_2}) \dots (j_{\nu-1}\vert u_{j_{\nu}})\rbra{j_\nu}\,
\end{align}
The sum consists of products of off-diagonal matrix elements indexed by sets $\{j_1,\dots,j_\nu\}$. These indices can be interpreted as nodes of a path and the off-diagonal matrix elements $(\ell\vert u_k)$ act as  propagators determining the amplitude of jumping $k-\ell$ steps inside the light cone. Crucially, these only depend on the difference $k-\ell$ and the amplitudes for negative step sizes vanish, making sure only causal paths contribute to the OTOC [Fig.~\ref{fig:circuits}(b)]. From Eq.~\eqref{eq:matrixelements} it follows that jumps of size $k$ are controlled by $z_k/q^k$, such that jumps with a large $k$ are exponentially suppressed.

We note that having maximal operator entanglement implies that all off-diagonal matrix elements vanish. Hence, in dual-unitary circuits the edges of an operator string can only move along the edges of the light cone and we recover the previous result. 

Although the mathematical origin is different, the picture presented here is qualitatively similar to the one presented in Ref.~\cite{kos2021correlations} for the two-point functions. The skeleton diagrams appearing there resemble the scattering paths introduced above. The similarity of the two results might merely be a reflection of the underlying physics of dual-unitary circuits. In dual-unitary circuits all excitations move with maximal velocity. Breaking dual unitarity then allows processes that violate this rule, suggesting expansions in orders of such processes.

\begin{figure}[tb]
    \centering
    \includegraphics[width = 0.45\textwidth]{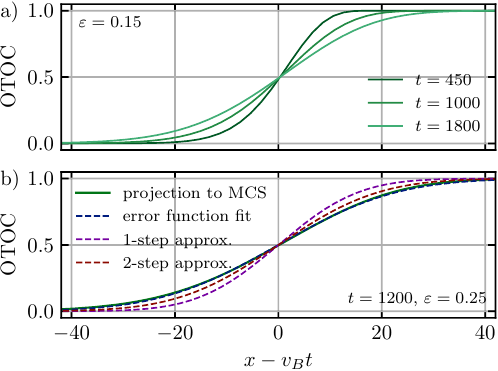}
    \caption{(a) Broadening of the OTOC front as a function of time. (b) Comparison of the asymptotic front for different approximations (the shift in the center of the front is corrected for better comparability).}
    \label{fig:front}
    \vspace{-\baselineskip}
\end{figure}

\emph{Truncation of the path integral.} 
To make analytical progress, we restrict the sum over paths. In the large-$q$ limit the path integral is dominated by those paths in which single steps are at most of size 1, since higher-order steps are suppressed by powers of $q$ [see Eq.~\eqref{eq:matrixelements}].

We now approximate the OTOC for general $q$ by only considering these paths, and call this the one-step approximation. On the level of scattering amplitudes this approximation corresponds to letting only $z_1$ be nonzero. However, even for qubits ($q=2$) the one-step approximation already produces the correct functional form of the asymptotic profile close to the center of the front [Fig.~\ref{fig:front}(c)]. 

Within this approximation the path integral can be evaluated exactly, leading to 
\begin{subequations}
\begin{align}
    C^{+}_{(1)} &= \frac{q^2F_{z_1}(n,m)-F_{z_1}(n-1,m)}{q^2-1} \label{eq:front_d1}, 
    \\ F_{z_1}(n,m)&\equiv n \binom{m}{n}B_{z_1}(n,m-n+1),  \label{eq:fz}
\end{align}
\end{subequations}
where $B_{z_1}(a,b)$ denotes the incomplete $\beta$-function. An asymptotic expansion yields a butterfly velocity $v_{B,1}=(1-z_1)/(1+z_1)$ and a front that takes the form
\begin{equation}
    C^{+}_{(1)} \approx \frac{1}{2}\left(1+\operatorname{erf}\left(\frac{x-v_{B,1}t}{\sqrt{2D_1t}}\right)\right). \label{eq:profile}
\end{equation}
This is the form expected from a diffusively broadening front with diffusion constant  $D_1=v_{B,1}(1-v_{B,1}^2)$. Both the form of the front and the scaling of the diffusion constant for $v_{B,1}\rightarrow1$ agree with results from Haar random circuits~\cite{von2018operator,nahum2018operator}. 
\begin{figure}[tb]
    \centering
    \includegraphics[width = 0.45\textwidth]{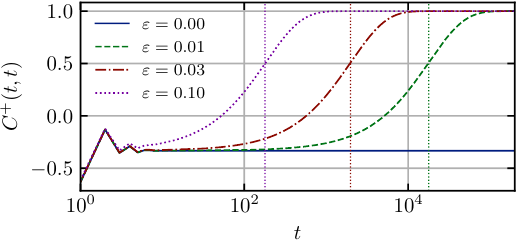}
    \caption{Early time relaxation of the dual-unitary form to the generic shape on the geometric light cone for a range of perturbation strengths. The timescale is indicated by vertical dotted lines. }
    \label{fig:early}
\end{figure}
\begin{figure}[tb]
    \centering
    \includegraphics[width = 0.45\textwidth]{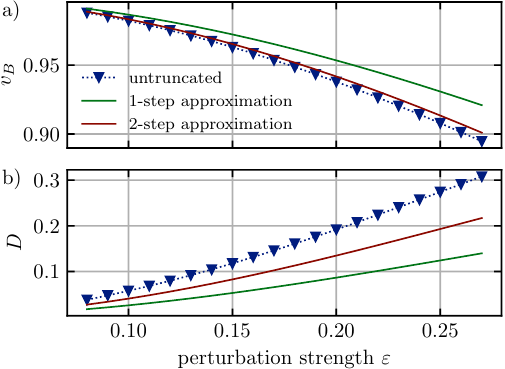}
    \caption{Plots of (a) butterfly velocity and (b) diffusion constant as function of perturbation strength and comparison to the analytical predictions of the one- and two-step approximations.}
    \label{fig:plots}
    \vspace{-\baselineskip}
\end{figure}

To obtain a better quantitative agreement for $q=2$, we consider the two-step approximation. An exact calculation yields \cite{SuppMat}
\begin{equation}
    C^{+}_{(2)}(x,t) \approx C^{+}_{(1)}\left(x+\frac{t-x}{2}\xi,t-\frac{t-x}{2}\xi\right),
\end{equation}
with $\xi=z_2/(q^2z_1)$. This result can be understood by noting that the path integral is asymptotically dominated by the typical path. If $z_2\ll z_1$ the typical fraction of steps of size two is approximately given by the ratio of scattering amplitudes $\xi=z_2/(q^2z_1)$. The steps of size two therefore effectively shift the profile [Eq.~\eqref{eq:fz}] deeper into the light cone, $n\rightarrow n(1-\xi)$. 

Close to the shifted front the shape of the profile is preserved,  with renormalized parameters
%\begin{subequations}
    \begin{align}
         v_{B,2} = \frac{v_{B,1}-\delta}{1-\delta},\,\,
        D_2 = D_1\frac{1-(1-v_{B,2})\frac{\xi}{2}}{\left(1-\delta\right)^2},
    \end{align}
%\end{subequations}
where $\delta=(1+v_{B,1})\xi/2$. By taking into account longer steps, operator strings can move into the light cone faster,  diminishing the butterfly velocity and, since this increases the variance of the distribution of the endpoints of operators, enhancing diffusion.

\emph{Numerical results.}  
Restricting the dynamics to the MCS (of linear size in $n$) presents an enormous simplification compared to the full exponentially large space when probing the long time limit. This restriction allows the efficient numerical evaluation of the OTOC under the assumptions stated above. 

First, we study the behavior of the OTOC on the geometric light cone for a Floquet circuit consisting of identical perturbed dual-unitary gates with $q=2$. We find that the main features of  dual unitarity persist up to the timescale~\eqref{eq:timescale} [Fig.~\ref{fig:early}]. At this timescale, the deviation of the OTOC on the geometric light cone $C^+(t,t)$ from the dual-unitary prediction, $-1/(q^2-1)=-1/3$, becomes of order $1$. This indicates that for earlier times most of the operator strings still travel at maximal velocity. 

In the late-time regime the operator front slows down, moving ballistically with $v_B<1$, and shows approximate diffusive broadening [see Fig.~\ref{fig:front}(a,b)]. At late times the shape of the operator front is well described by an error function of the form described in Eq.~\eqref{eq:profile}, indicating that higher $k$-steps only serve to further renormalize the arguments of the obtained profile.

We extract the butterfly velocity and diffusion constant and compare them to the analytic prediction obtained by truncating the path integral [Fig.~\ref{fig:plots}]. For a specific but randomly selected  perturbation we find that the two-step approximation is in good agreement with the full path integral result for low to intermediate perturbation strengths. For the diffusion constant the discrepancy is larger, but the qualitative behavior is captured. We attribute this discrepancy to paths which contain large steps. 
For both quantities, the two-step approximation significantly improves the one-step approximation, and the accuracy of this approximation is expected to increase for larger $q$ or by including higher steps.

\emph{Discussion.}

The observation of diffusively broadening fronts in non-random systems far away from the dual-unitary limit~\cite{xu2020accessing,LopezPiqueres2021} hints at the presence of a more general mechanism.  Numerical results indicate that the degeneracy of the subleading eigenvalue can remain stable far from dual unitarity, suggesting that a similar path-integral description remains possible beyond the perturbative regime. Moreover, the non-Hermiticity of the transfer matrix might play a central role -- Ref.~\cite{Bensa2022} previously observed that non-Hermiticity can strongly influence the behavior of OTOCs. 

Our work shows that dual-unitary circuits can serve as a starting point to investigate more generic settings. We hope that this work opens up further studies on perturbed dual-unitary cicuits, e.g., on  entanglement dynamics, the relation to transport, or spectral properties. The developed framework can be directly applied to more general probes of operator dynamics in perturbed dual-unitary dynamics involving multiple replicas of the circuit, e.g. R\'enyi (operator) entanglement \cite{bertini2020operator,zhou2022maximal}, spectral form factors \cite{bertini2018exact,bertini2021random}, or in studies of `deep thermalization' \cite{Ho2022,Claeys2022,Ippoliti2022}.
\\

\begin{acknowledgments}
We are grateful to Dominik Hahn, Pavel Kos, Chris R. Laumann, David M. Long, Frank Pollmann, Toma\v{z} Prosen, and Philippe Suchsland for useful discussions. This work was in part funded by the Deutsche Forschungsgemeinschaft (DFG) via the cluster of excellence ct.qmat (EXC 2147, project-id 390858490).
\end{acknowledgments}

\bibliography{perturbed_DU}% Produces the bibliography via BibTeX.

\setcounter{equation}{0}

\setcounter{figure}{0}

\setcounter{table}{0}

\setcounter{section}{0}

\makeatletter

\renewcommand{\theequation}{S\arabic{equation}}

\renewcommand{\thefigure}{S\arabic{figure}}

\renewcommand{\thesection}{S-\Roman{section}}

%\renewcommand{\bibnumfmt}[1]{[S#1]}

%\renewcommand{\citenumfont}[1]{S#1} 

%%%%%%%%%% Prefix a "S" to all equations, figures, tables and reset the counter %%%%%%%%%%
\begin{widetext}
\clearpage
\begin{center}
\textbf{Supplemental Material: \\ From Dual Unitarity to Generic Quantum Operator Spreading}\\
Michael A. Rampp, Roderich Moessner, and Pieter W. Claeys\\
\textit{Max Planck Institute for the Physics of Complex Systems, 01187 Dresden, Germany}\\
\maketitle
%\vspace{-\baselineskip}
\end{center}

In this supplemental material we detail the calculations mentioned in the main text. The first section reviews the construction and properties of the maximally chaotic subspace (MCS), as well as its application to the computation of out-of-time-ordered correlators (OTOCs) in dual-unitary circuits. Next, the matrix elements of general gates in the MCS are computed. In the second section the quantum-information-theoretic interpretation of the scattering amplitudes $z_k$ is elaborated upon and used to prove various bounds. The third section presents details of the derivation of the discrete path integral formula and the profile of the OTOC in the $1$- and two-step approximation is derived and its asymptotic form derived. 

\section{Transfer Matrix in the Maximally Chaotic Subspace}

In this section the construction of the MCS and the computation of OTOCs in dual-unitary circuits is reviewed (for a more detailed derivation see Ref.~\cite{claeys2020maximum}), and the matrix elements of the transfer matrix for perturbed circuits in this space are computed. For an overview of unitary circuits, see Ref.~\cite{Fisher2022}. The conventions used in the following are introduced in the main text. 

\subsection{Construction of the Maximally Chaotic Subspace}

As introduced in the main text, the OTOC is defined as
\begin{align}
C_{\alpha\beta}(x,t) =  \langle \sigma_{\alpha}(0,t)\sigma_\beta(x,0)\sigma_{\alpha}(0,t)\sigma_\beta(x,0)\rangle. \label{eq:otoc_sm}
\end{align}
Following Ref.~\cite{claeys2020maximum}, after representing this equation graphically and eliminating all gates outside the causal light cones of $\sigma_{\alpha}$ and $\sigma_{\beta}$, the OTOC can be represented as the following tensor network (TN) contraction for $t-x\in2\mathbb{Z}$
\begin{align}
C^+(n,m) &= \frac{1}{q^{n+m}}\,\vcenter{\hbox{\includegraphics[width=0.35\columnwidth]{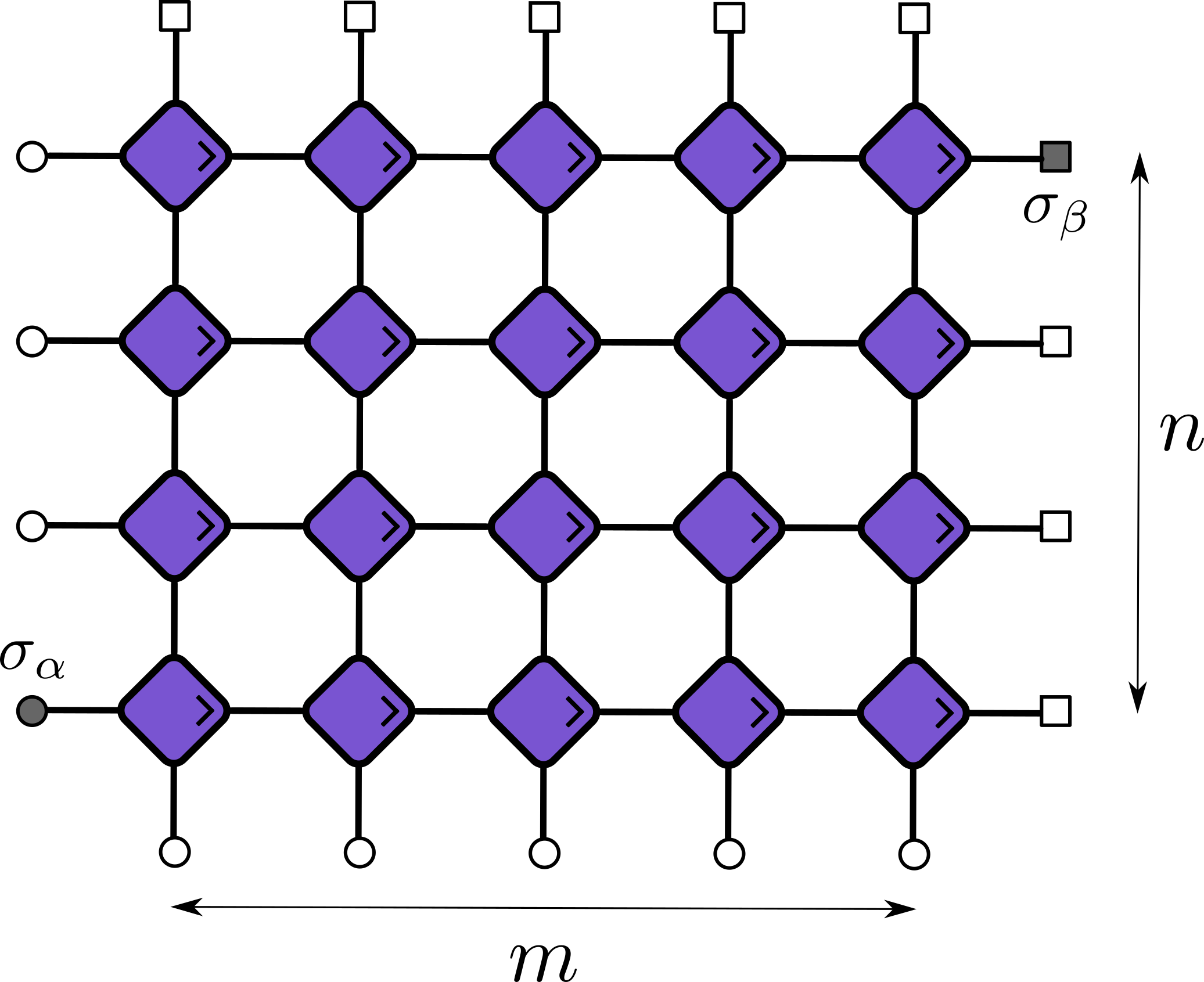}}} \label{fig:otoc_plus_sm},
\end{align}
where the light-cone coordinates $n=(t-x+2)/2,\,m=(t+x)/2$ set the size of the TN. For $t-x\in2\mathbb{Z}+1$ we find
\begin{align}
C^-(n,m) &= \frac{1}{q^{n+m}}\,\vcenter{\hbox{\includegraphics[width=0.35\columnwidth]{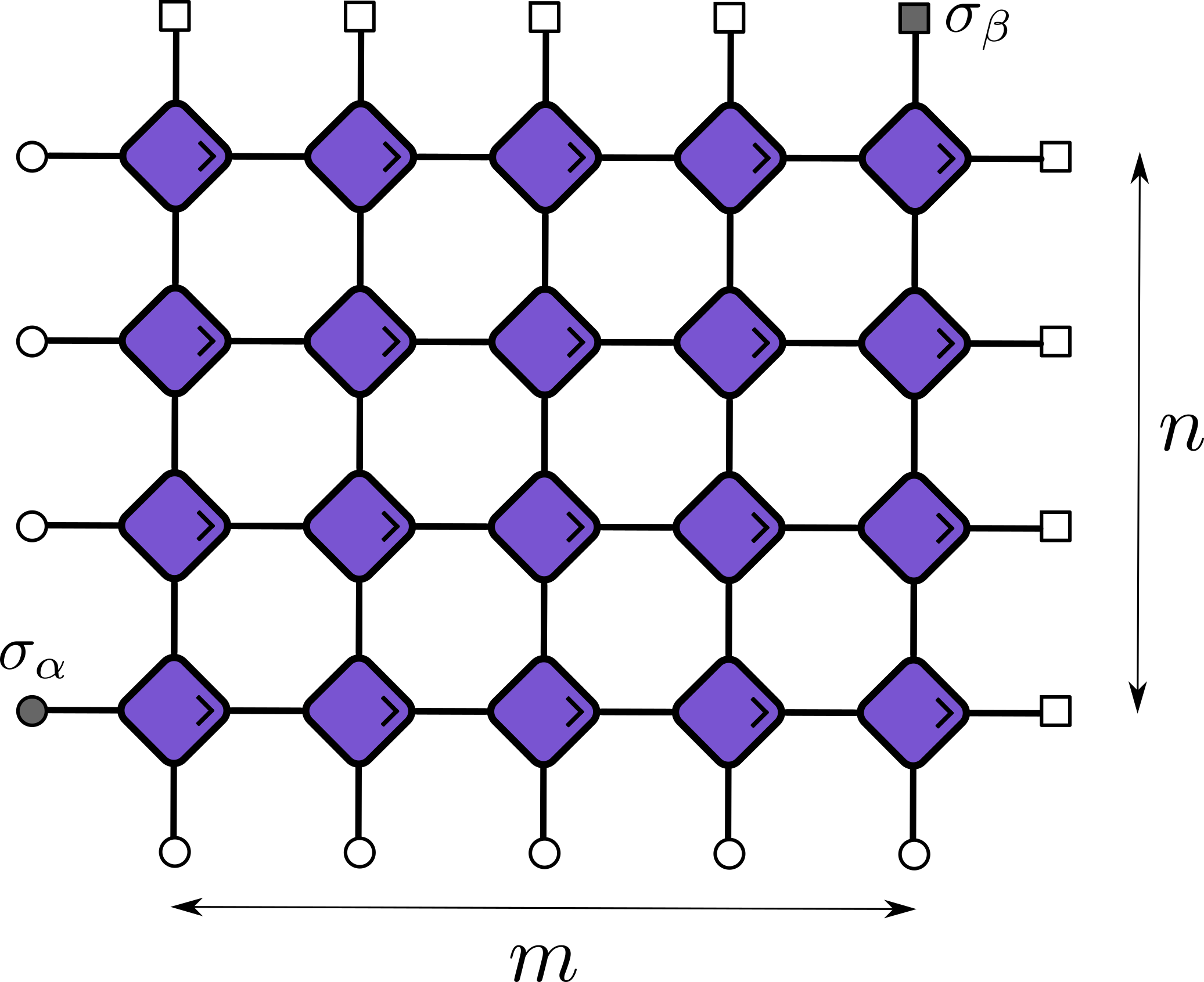}}} \label{fig:otoc_minus},
\end{align}
and in this case $n=(t-x+1)/2,\,m=(t+x+1)/2$.  

We are interested in the dynamics at long times and relatively close to the geometric light cone $t=x$. Therefore, we consider the TN in the limit of large $m$ as being generated by the \emph{column transfer matrix} $T_n$, defined as
\begin{align}
T_n = \frac{1}{q}\,\vcenter{\hbox{\includegraphics[width=0.13\columnwidth]{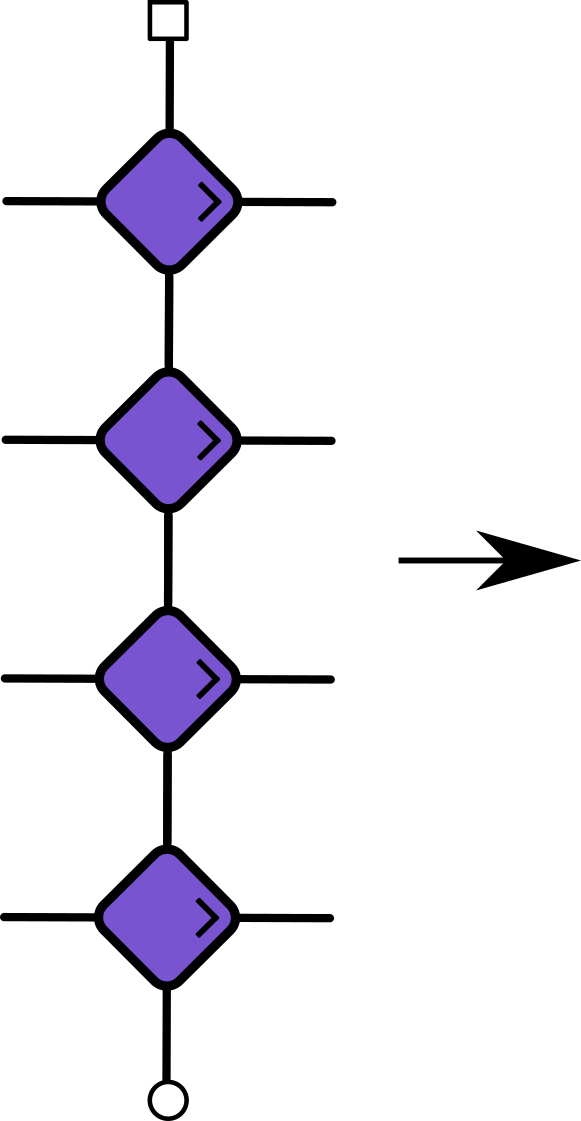}}}\quad\frac{1}{q}\,\vcenter{\hbox{\includegraphics[width=0.27\columnwidth, angle=180]{figs/transfer_mat_rotated.png}}} \label{fig:tm}\,,
\end{align}
which we rotate by $90^\circ$ for convenience. We introduce the vectors corresponding to the left and right boundary conditions 
\begin{equation}
    \rbra{L_n(\sigma_\alpha)} = \frac{1}{q^{n/2}}\,\vcenter{\hbox{\includegraphics[width=0.03\textheight,angle=180,trim={0 0.45cm 0 0},clip]{figs/circle_filled.png}}}\,\vcenter{\hbox{\includegraphics[width=0.015\textheight,angle=180]{figs/circle.png}}}\dots\vcenter{\hbox{\includegraphics[width=0.015\textheight,angle=180]{figs/circle.png}}},\qquad \rket{R_n^+(\sigma_\beta)} = \frac{1}{q^{n/2}}\,\vcenter{\hbox{\includegraphics[width=0.015\textheight,]{figs/square.png}}}\dots\vcenter{\hbox{\includegraphics[width=0.015\textheight]{figs/square.png}}}\vcenter{\hbox{\includegraphics[width=0.03\textheight,trim={0 0.45cm 0 0},clip]{figs/square_filled.png}}}.
\end{equation}
For odd $(t-x)$ the right boundary condition reads
\begin{equation}
    \rket{R_n^-(\sigma_\beta)} = \frac{1}{q^{n/2}} \,\vcenter{\hbox{\includegraphics[height=0.2\textheight,angle=-90]{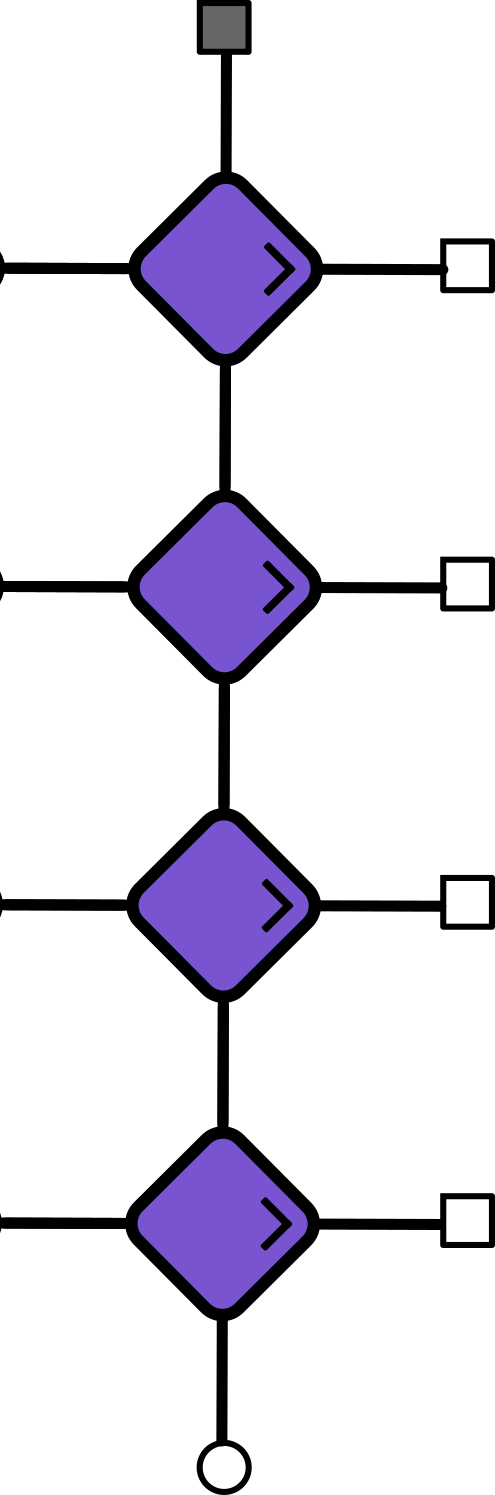}}}\,,
\end{equation}
such that we can write
\begin{equation}
    C^+_{\alpha\beta}(n,m) =  \rbra{L_n(\sigma_\alpha)}(T_n)^m\rket{R_n^+(\sigma_\beta)}, \quad C^-_{\alpha\beta}(n,m) =  \rbra{L_n(\sigma_\alpha)}(T_n)^{m-1}\rket{R_n^-(\sigma_\beta)}.
\end{equation}

The transfer matrix is a contracting map, i.e. $\| T_n v\|\leq\|v\|$. Hence, all its eigenvalues lie inside or on the boundary of the complex unit disk. In the limit $m\rightarrow\infty$ the bulk of the TN becomes a projector onto the leading (modulus $1$)  eigenspace of the transfer matrix. For circuits composed of generic unitary gates the leading eigenspace is nondegenerate. As $T_n$ is in general not symmetric the leading left- and right-eigenvectors are not related by transposition. In the folded representation unitarity is expressed as
\begin{align}
    \vcenter{\hbox{\includegraphics[width=0.08\columnwidth]{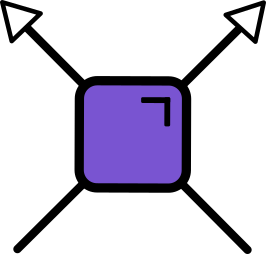}}}\, = \vcenter{\hbox{\includegraphics[width=0.012\textheight,angle=225]{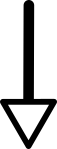}}}\,\vcenter{\hbox{\includegraphics[width=0.012\textheight,angle=135]{figs/triangle.png}}}, \qquad
    \vcenter{\hbox{\includegraphics[width=0.08\columnwidth]{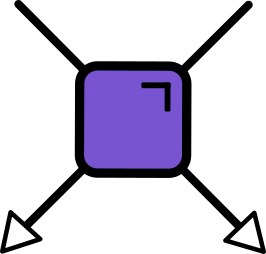}}}\, = \vcenter{\hbox{\includegraphics[width=0.012\textheight,angle=-45]{figs/triangle.png}}}\,\vcenter{\hbox{\includegraphics[width=0.012\textheight,angle=45]{figs/triangle.png}}},
\end{align}
where the triangle denotes an arbitrary permutation of even and odd legs. These relations imply that the following two vectors are leading eigenvectors
\begin{align}
\rket{n,0} = \frac{1}{q^n}\,\overbrace{\vcenter{\hbox{\includegraphics[width=0.012\textheight]{figs/square.png}}}\,\,\vcenter{\hbox{\includegraphics[width=0.012\textheight]{figs/square.png}}}\,\,\dots\,\,\vcenter{\hbox{\includegraphics[width=0.012\textheight]{figs/square.png}}}\,\,\vcenter{\hbox{\includegraphics[width=0.012\textheight]{figs/square.png}}}}^{n}, \qquad
\rbra{n,n} = \frac{1}{q^n}\,\overbrace{\vcenter{\hbox{\includegraphics[width=0.012\textheight,angle=180]{figs/circle.png}}}\,\,\vcenter{\hbox{\includegraphics[width=0.012\textheight,angle=180]{figs/circle.png}}}\,\,\dots\,\,\vcenter{\hbox{\includegraphics[width=0.012\textheight,angle=180]{figs/circle.png}}}\,\,\vcenter{\hbox{\includegraphics[width=0.012\textheight,angle=180]{figs/circle.png}}}}^{n}.
\end{align}
Dual unitarity, which is expressed in the folded language as,
\begin{equation}
    \vcenter{\hbox{\includegraphics[width=0.08\columnwidth]{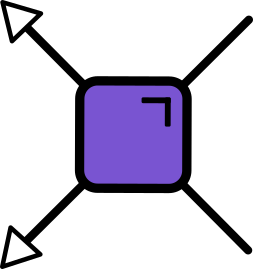}}}\, = \begin{array}{c}
\vcenter{\hbox{\includegraphics[width=0.012\textheight,angle=-135]{figs/triangle.png}}}\\
\vcenter{\hbox{\includegraphics[width=0.012\textheight,angle=-45]{figs/triangle.png}}}
\end{array}, \qquad
\vcenter{\hbox{\includegraphics[width=0.08\columnwidth]{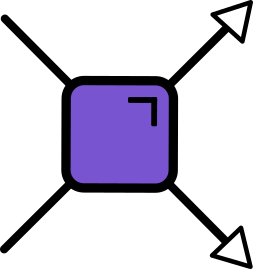}}}\, = \begin{array}{c}
\vcenter{\hbox{\includegraphics[width=0.012\textheight,angle=135]{figs/triangle.png}}}\\
\vcenter{\hbox{\includegraphics[width=0.012\textheight,angle=45]{figs/triangle.png}}}
\end{array},
\end{equation}
allows the construction of further leading eigenvectors. Using the above identities, it can be shown that vectors of the form
\begin{align}
    \rket{n,k} = \frac{1}{q^n}\,\overbrace{\vcenter{\hbox{\includegraphics[width=0.012\textheight]{figs/circle.png}}}\,\,\dots\,\,\vcenter{\hbox{\includegraphics[width=0.012\textheight]{figs/circle.png}}}}^{k}\,\,\overbrace{\vcenter{\hbox{\includegraphics[width=0.012\textheight]{figs/square.png}}}\,\,\dots\,\,\vcenter{\hbox{\includegraphics[width=0.012\textheight]{figs/square.png}}}}^{n-k}
\end{align}
are right eigenvectors with eigenvalue 1. Moreover, the associated transposed vectors $\rbra{n,k}$ are left eigenvectors with eigenvalue 1. These vectors can be used to construct a set of orthonormal eigenvectors as
\begin{equation}
    \rket{\overline{n,0}} := \rket{n,0}, \qquad \rket{\overline{n,k}} := \frac{q\rket{n,k}-\rket{n,k-1}}{\sqrt{q^2-1}},\quad k\geq1.
\end{equation}\,
We call the space spanned by these vectors the maximally chaotic subspace (MCS)~\cite{bertini2020operator}. In the main text we adopt the simplified notation $\rket{k}:=\rket{\overline{n,k}}$ for convenience. The MCS is constructed without any reference to the properties of the particular gate, apart from dual unitarity. If the leading eigenspace of the column transfer matrix constructed from a gate equals the MCS, this gate is called \emph{maximally chaotic}. Not every dual-unitary gate is maximally chaotic, in particular the existence of a local conserved quantity always leads to additional leading eigenvectors, but the set of maximally chaotic gates is dense in the set of dual-unitary gates.

\subsection{OTOCs in Maximally Chaotic Dual-Unitary Circuits}

For maximally chaotic gates the OTOC acquires a universal form in the late-time regime. As discussed above, for $m\rightarrow\infty$ the bulk of the TN becomes a projector on the leading eigenspace, in this case the MCS. Thus, to compute the OTOC it only remains to contract the vectors in the MCS with the boundary conditions of the TN.
The nonzero overlaps are given by
\begin{subequations}
    \begin{align}
        &\left(L_n\vert\overline{n,0}\right) =\frac{1}{q^{n/2}},  \qquad \left(L_n\vert\overline{n,1}\right) = -\frac{1}{q^{n/2}\sqrt{q^2-1}}, \qquad
        \left(\overline{n,n}\vert R_n^+\right) = \frac{q}{q^{n/2}\sqrt{q^2-1}},\\
        &\left(\overline{n,k}\vert R_n^-\right) = \frac{q^{n/2+1}}{q^{k}\sqrt{q^2-1}}\left(\mathcal{M}_{n-k}\left(\sigma_\beta\right)-\mathcal{M}_{n-k+1}\left(\sigma_\beta\right)\right).
    \end{align}
\end{subequations}
Here, we have introduced the map $\mathcal{M}_n(\sigma_\beta):=\tr{\mathcal{M}_+^n(\sigma_\beta)^\dagger \mathcal{M}_+^n(\sigma_\beta)}/q$ in terms of the light-cone channel $\mathcal{M}_+$,
\begin{align}
    \mathcal{M}_+(\sigma) := \frac{1}{q}\,\,\vcenter{\hbox{\includegraphics[height=0.12\textwidth]{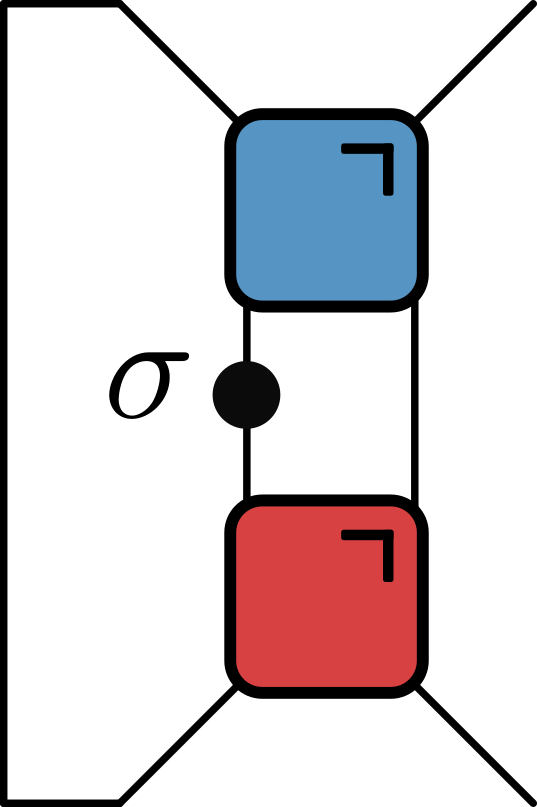}}}.
\end{align}
Hence, replacing $T_n^m$ by a projector in the MCS can be used to evaluate the long-time value of the OTOC in generic dual-unitary circuits as
\begin{subequations}
\begin{align}
    \lim_{m\rightarrow\infty} C^+_{\alpha\beta}(n,m) &= \sum_{k=0}^n \left(L_n(\sigma_\alpha)\vert\overline{n,k}\right) \left(\overline{n,k}\vert R_n^+(\sigma_\beta)\right) = -\frac{1}{q^2-1}\delta_{n,1},\\
    \lim_{m\rightarrow\infty} C^-_{\alpha\beta}(n,m) &= \sum_{k=0}^n \left(L_n(\sigma_\alpha)\vert\overline{n,k}\right) \left(\overline{n,k}\vert R_n^-(\sigma_\beta)\right) = \frac{q^2\mathcal{M}_{n}\left(\sigma_\beta\right)-\mathcal{M}_{n-1}\left(\sigma_\beta\right)}{q^2-1}.
\end{align}
\end{subequations}

\subsection{Projected Transfer Matrix for General Circuits}

We perform degenerate perturbation theory in the MCS. For this we need to compute the matrix elements of the column transfer matrix in the MCS. First of all we compute $\rbra{n,\ell}T_n\rket{n,k}$ for $\ell\geq k$ and find the same result as for dual-unitary gates
\begin{align}
    \rbra{n,\ell}T_n\rket{n,k} = \frac{1}{q^{2n+1}}
        \aoverbrace[*{4}{0}L1U1R*{4}{0}]{
         \vcenter{\hbox{\includegraphics[height=0.06\textheight]{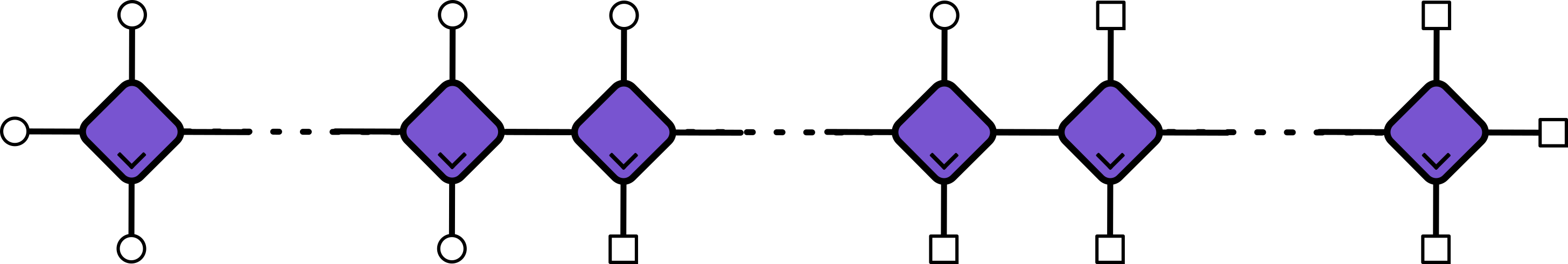}}}}^{\ell-k}
     = \frac{1}{q^{\ell-k}},
\end{align}
since the unitarity condition can be applied to every gate and we obtain a number that does not depend on any microscopic properties of the gates. Next we compute
\begin{align}
   \rbra{n,0}T_n\rket{n,k} =&  \frac{1}{q^{2n+1}}\aunderbrace[*{1}{0}l1D1r*{5}{0}]{\vcenter{\hbox{\includegraphics[height=0.06\textheight]{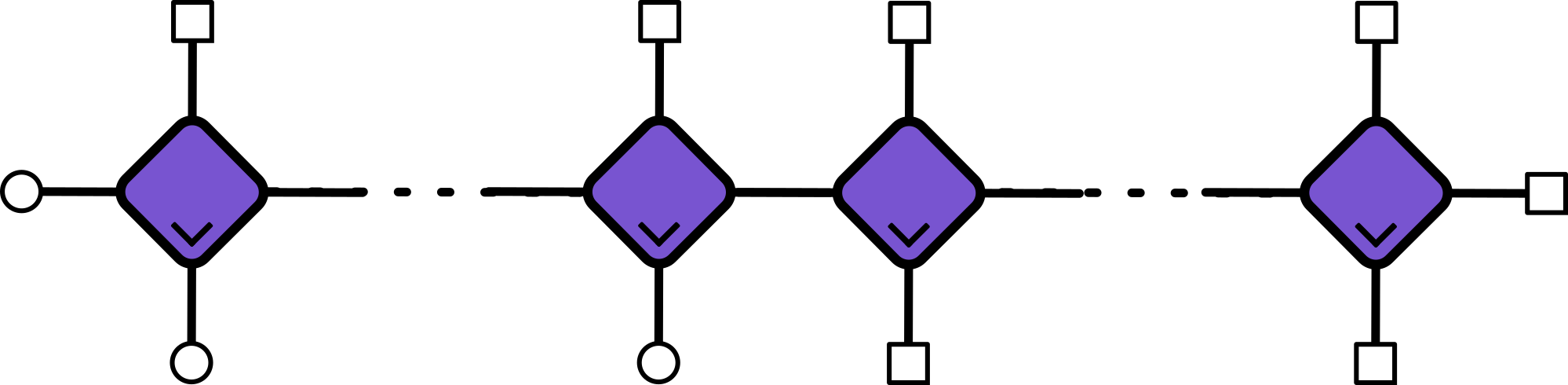}}}}_{k}= \frac{1}{q^{2k+1}}\vcenter{\hbox{\includegraphics[height=0.07\textheight,trim={0 1cm 0 0},clip]{figs/gen_bubble_reduced.png}}}
   = \frac{B_k}{q^{k}},
\end{align}
where we have used the unitarity of the gate and expressed the matrix element through $B_k$ as introduced in the main text. Finally, for $0<\ell<k$ we find
\begin{align}
    \rbra{n,\ell}T_n\rket{n,k} =&  \frac{1}{q^{2n+1}}\vcenter{\hbox{\includegraphics[height=0.06\textheight]{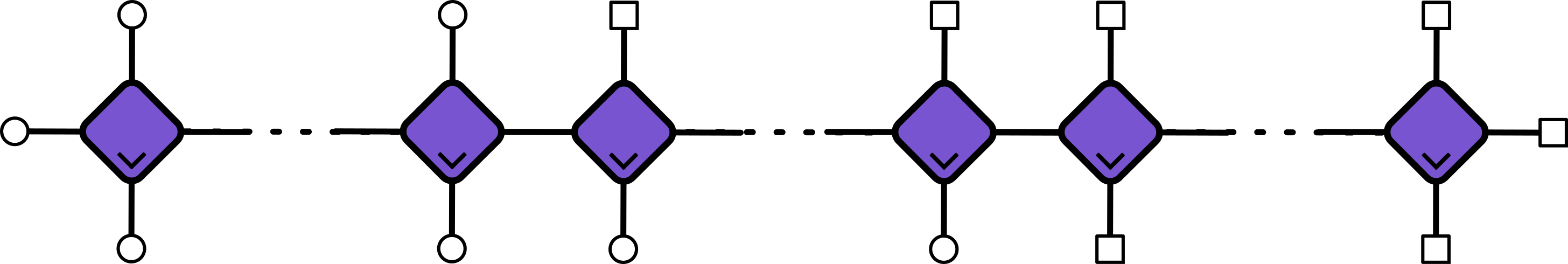}}}\nonumber\\
   =& \frac{1}{q^{2(k-\ell)+1}}\underbrace{\vcenter{\hbox{\includegraphics[height=0.07\textheight,trim={0 1cm 0 0},clip]{figs/gen_bubble_reduced.png}}}}_{k-\ell}
   = \frac{B_{k-\ell}}{q^{k-\ell}}.
\end{align}
Translating these results into the orthonormal basis $\{\rket{\overline{n,k}}\}$ yields
\begin{subequations}
\label{eq:matrixelements_sm}
    \begin{align}
    \rbra{\overline{n,0}}T_n\rket{\overline{n,0}} &= 1\\
    \rbra{\overline{n,0}} T_n\rket{\overline{n,k}} &= \sqrt{q^2 -1}\frac{z_k}{q^{k-1}}, \quad k\geq1 \\
    \rbra{\overline{n,\ell}} T_n\rket{\overline{n,k}} &= \begin{cases} \frac{q^2 z_{k-l} - z_{k-l+1}}{q^{k-l}}, &\,\, k>l, \\ 1-z_1, &\,\, k=l, \\ 0, &\,\, \mathrm{otherwise}. \end{cases}
    \end{align}
\end{subequations}
The quantities $z_k=(B_k-B_{k-1})/(q^2-1)$ have been introduced in the main text.

Notice that the transfer matrix in the MCS has a non-trivial Jordan structure. The algebraic multiplicity of $\lambda_1=1-z_1$ is $n$, while its geometric multiplicity is $1$. The Jordan structure is crucial in obtaining a diffusive OTOC profile, since a diagonalizable transfer matrix can only give rise to exponential decay (given the overlaps of  MCS vectors with the boundary conditions).

\subsection{Dilute Limit}

\begin{figure}[tb]
    \centering
    \includegraphics[width = 0.45\textwidth]{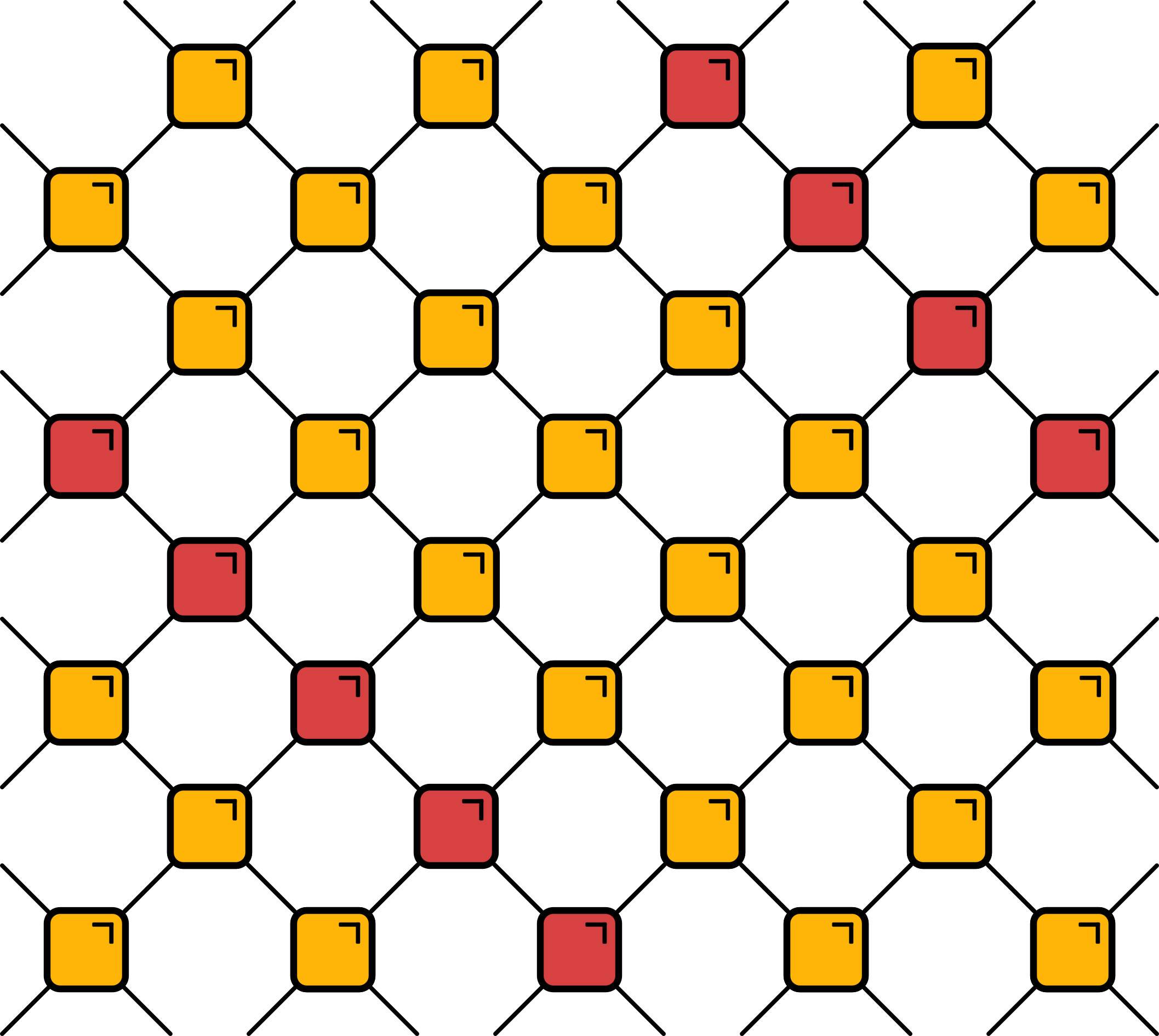}
    \caption{Time-evolution operator for a circuit composed of lines of defects orthogonal to the light cone (marked in red) with patches of maximally chaotic dual unitary gates in between (marked in yellow). In the limit of large separation between the lines of defects the projection to the MCS is controlled.}
    \label{fig:dilute_sm}
    \vspace{-\baselineskip}
\end{figure}

In the following we argue that the projection of the transfer matrix to the MCS is controlled in a particular dilute limit.
Note that after repeated application the transfer matrix of a maximally chaotic dual-unitary circuit approaches the projector on the MCS, $\lim_{m\rightarrow\infty}T_{n,DU}^m=\Pi_{\mathrm{MCS}}$. Hence, a circuit composed of lines of defects orthogonal to the light cone with patches of maximally chaotic dual unitary gates inbetween can be expressed using such a projected transfer matrix, up to corrections exponentially small in the distance between the defects. This is depicted in Fig.~\ref{fig:dilute_sm} for finite sized patches. 

\section{Scattering Amplitudes and Operator Entanglements}

In this section relevant properties of the matrix elements $B_k$ and the resulting scattering amplitudes $z_k$ are collected. It is shown that the $B_k$ can be expressed as operator entanglements on an enlarged Hilbert space. This construction allows to prove the bound $1\leq B_k\leq q^2$ (implying $z_k\leq1$). Furthermore, it is shown that for gates that are not dual-unitary the $B_k$ converge to $q^2$ for $k\rightarrow\infty$ (and hence $z_k\rightarrow0$ for $k\rightarrow\infty$).

\subsection{Equivalence of $B_1$ and Operator Entanglement}

In the following it is shown that the quantity $B_1$ is equivalent to the operator entanglement of the gate. This equivalence implies the bound $1\leq B_1\leq q^2$ and that $B_1=1$ if and only if the underlying gate is dual unitary. We begin by reviewing the operator-to-state mapping for two-site gates: an operator $U$ acting on two sites can be mapped to a state $\ket{\psi_U}$ on a 4-site Hilbert space by 
\begin{equation}
    \ket{\psi_U} = (U_{AB}\otimes 1_{CD})\ket{\phi}_{AC}\ket{\phi}_{BD},
\end{equation}
here $\ket{\phi}=\frac{1}{\sqrt{q}}\sum_j\ket{j}\ket{j}$ denotes the Bell state on two sites. Operationally, two Bell pairs are created and subsequently entangled by the gate $U$. This is clearer in the graphical representation:
\begin{align}
\ket{\psi_U}=\frac{1}{q}\,\vcenter{\hbox{\includegraphics[width=0.14\columnwidth]{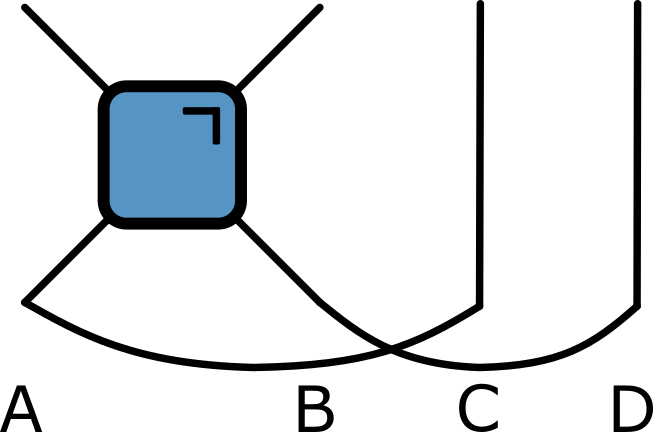}}} \label{eq:op_to_state}
\end{align}
It is useful to introduce the Schmidt decomposition of $U$, $U= \sum_{j=1}^{q^2}\sqrt{\sigma_j} X_j \otimes Y_j$, where $\sigma_j\geq0$ are the Schmidt coefficients, and $X_j$ and $Y_j$ denote orthonormal operator bases of the local Hilbert space, $\operatorname{tr}[X_j^\dagger X_k]=\delta_{jk}$ and $\operatorname{tr}[Y_j^\dagger Y_k]=\delta_{jk}$. This implies a decomposition of the state $\ket{\psi_U}$ as
\begin{equation}
    \ket{\psi_U} = \frac{1}{q} \sum_{j,k} \sqrt{\sigma_j\sigma_k} X_j\ket{j}_A\otimes Y_k\ket{k}_B \otimes \ket{j}_C \otimes \ket{k}_D.
\end{equation}
The entanglement measure that is related to $B_1$ is the purity of the reduced density matrix on the subset $AC$. It holds that
\begin{equation}
    \operatorname{Tr}[\rho_{AC}^2] = \frac{1}{q^{4}}\sum_{j=1}^{q^{2}}\sigma_j^2.
\end{equation}
The so-called linear operator entanglement is then defined as $E(U)=1-\operatorname{Tr}[\rho_{AC}^2]$~\cite{zanardi2001entanglement}. Its maximal value is $E(U)=1-\frac{1}{q^2}$ when all Schmidt values are equal. We call such operators \emph{maximally entangled}. An operator being maximally entangled is equivalent to it satisfying unitarity in the spatial direction~\cite{rather2020creating}. Hence, dual unitary gates are those gates which are unitary and maximally entangled. For separable operators of the form $U=qX_A\otimes Y_B$ the operator entanglement takes the minimal value $E(U)=0$.

On the other hand, inserting the Schmidt decomposition into the matrix element $B_1$ we find
\begin{equation}
    B_1 = \frac{1}{q^{2}}\sum_{j=1}^{q^{2}}\sigma_j^2 = q^2\operatorname{Tr}[\rho_{AC}^2] = q^2(1-E(U)).
\end{equation}
The equivalence can also be seen from inspection of the corresponding diagram
\begin{align}
\operatorname{Tr}[\rho_{AC}^2]=\frac{1}{q^4}\,\vcenter{\hbox{\includegraphics[width=0.4\columnwidth]{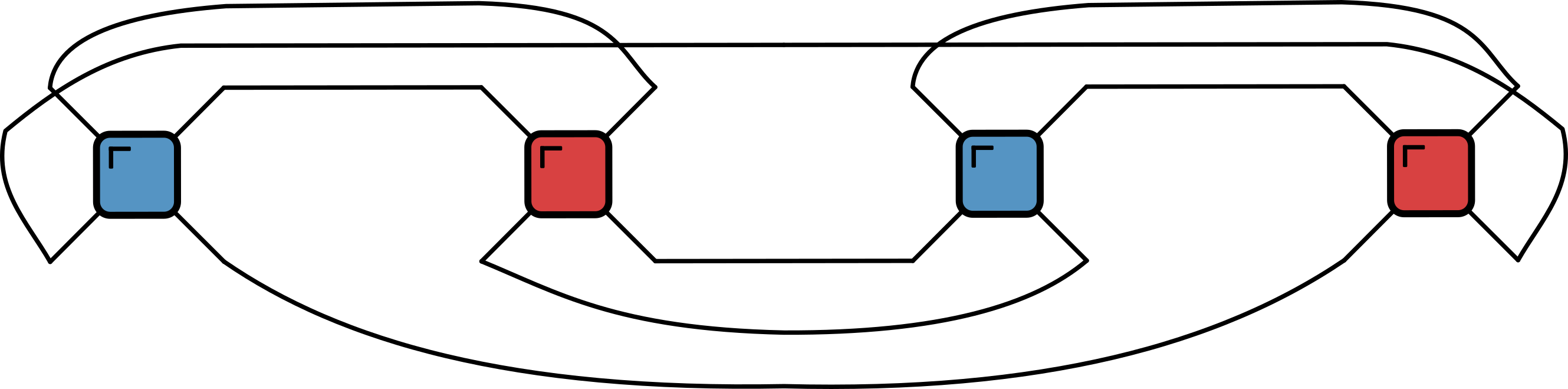}}}\,=\frac{1}{q^2} B_1 \label{eq:b1_trace}
\end{align}
The bound on $E(U)$ immediately implies $1\leq B_1\leq q^2$ and that $B_1=1$ is attained iff the gate is dual unitary.

\subsection{$B_k$ and Operator Entanglement of Diagonal Compositions}

In the following we generalize the above arguments to $B_k$ with $k>1$. We show that these matrix elements can also be expressed through the operator entanglement of a gate acting on a larger Hilbert space with respect to a particular bipartition. This immediately implies $1\leq B_k\leq q^{k+1}$, a generalized version of the bound derived in the previous section. We then use that the enlarged gate is constructed via diagonal composition to prove the more stringent bound $B_k\leq q^2$.

%In the following we show that the higher order matrix elements can be given a similar interpretation in terms of operator entanglement of a $k$-fold diagonally composed gate acting on $\mathbb{C}^{q}\otimes\mathbb{C}^{q^k}$. 
Consider the following Schmidt decomposition of  a gate $\mathfrak{U}\in U(q^{k+1})$
\begin{equation}
    \mathfrak{U} = \sum_{j=1}^{q^{k+1}} \sqrt{\sigma_j} X_j \otimes Y_j, \label{eq:schmidt}
\end{equation}
with $X_j:\mathbb{C}^{q}\rightarrow\mathbb{C}^{q^k}$ and $Y_j:\mathbb{C}^{q^k}\rightarrow\mathbb{C}^{q}$ forming complete orthonormal bases. Using the graphical calculus we can show that $B_k$ can be expressed through these Schmidt values as
\begin{equation}
    B_k = \frac{1}{q^{k+1}} \sum_{j=1}^{q^{k+1}} \sigma_j^2.
\end{equation}
Notice that this result does not rely on %the fact that the gate has been obtained as a diagonal composition. 
any internal structure of $\mathfrak{U}$. We can define the operator entanglement with respect to the above partition as
\begin{equation}
    E_k(\mathfrak{U}) := 1 - \frac{1}{q^{2(k+1)}}\sum_{j=1}^{q^{k+1}} \sigma_j^2 \in [0,1-\frac{1}{q^{k+1}}]. \label{eq:higher_entanglement}
\end{equation}
This quantity determines the higher-order values of $B_k$ as
\begin{equation}
    B_k = q^{k+1} (1 - E_k(\mathfrak{U})) \in [1, q^{k+1}].
\end{equation}

In the following we make use of the fact that the gate $\mathfrak{U}$ determining $B_k$ is constructed out of the two-site gate $U$ by diagonal composition (see Eq.~\eqref{eq:op_to_state_general})  to derive the stronger bound $B_k\leq q^2$. We use an argument based on monogamy of entanglement. On the level of operators it expresses that operators constructed by diagonal composition cannot be separable with respect to the partition introduced above.

Formally, we introduce $2k+2$ sites with a local $q$-dimensional Hilbert space and consider the partitions 
\begin{subequations}
\begin{align}
\begin{array}{llll}
    A' =\{1\},& B' = \{2,...,k+1\},& C' = \{k+2\},& D' = \{k+3,...,2k+2\},\\
    A\textcolor{white}{'} =\{1,...,k\},& B\textcolor{white}{'} = \{k+1\},&  C\textcolor{white}{'} = C',&  D\textcolor{white}{'} = D'.
\end{array}
\end{align}
\end{subequations}
Then we define the state
\begin{equation}
    \ket{\psi_U} = (U_{k,k+1}\otimes \mathbbm{1}_{(k,k+1)^c})\dots(U_{1,2}\otimes \mathbbm{1}_{(1,2)^c})\ket{\phi}_{A'C'}\ket{\phi}_{B'D'},
\end{equation}
where $\ket{\phi}_{XY}$ denotes the generalized Bell state on the appropriate subspaces. In words, we prepare the subsystems $A'C'$ and $B'D'$ in maximally entangled states respectively and then we apply the diagonally composed unitary transformation $\mathfrak{U}$ to the subset $A'B'$. For illustration, in the case of $k=2$ we have the state
\begin{align}
\ket{\psi_U} =\,\vcenter{\hbox{\includegraphics[width=0.2\columnwidth]{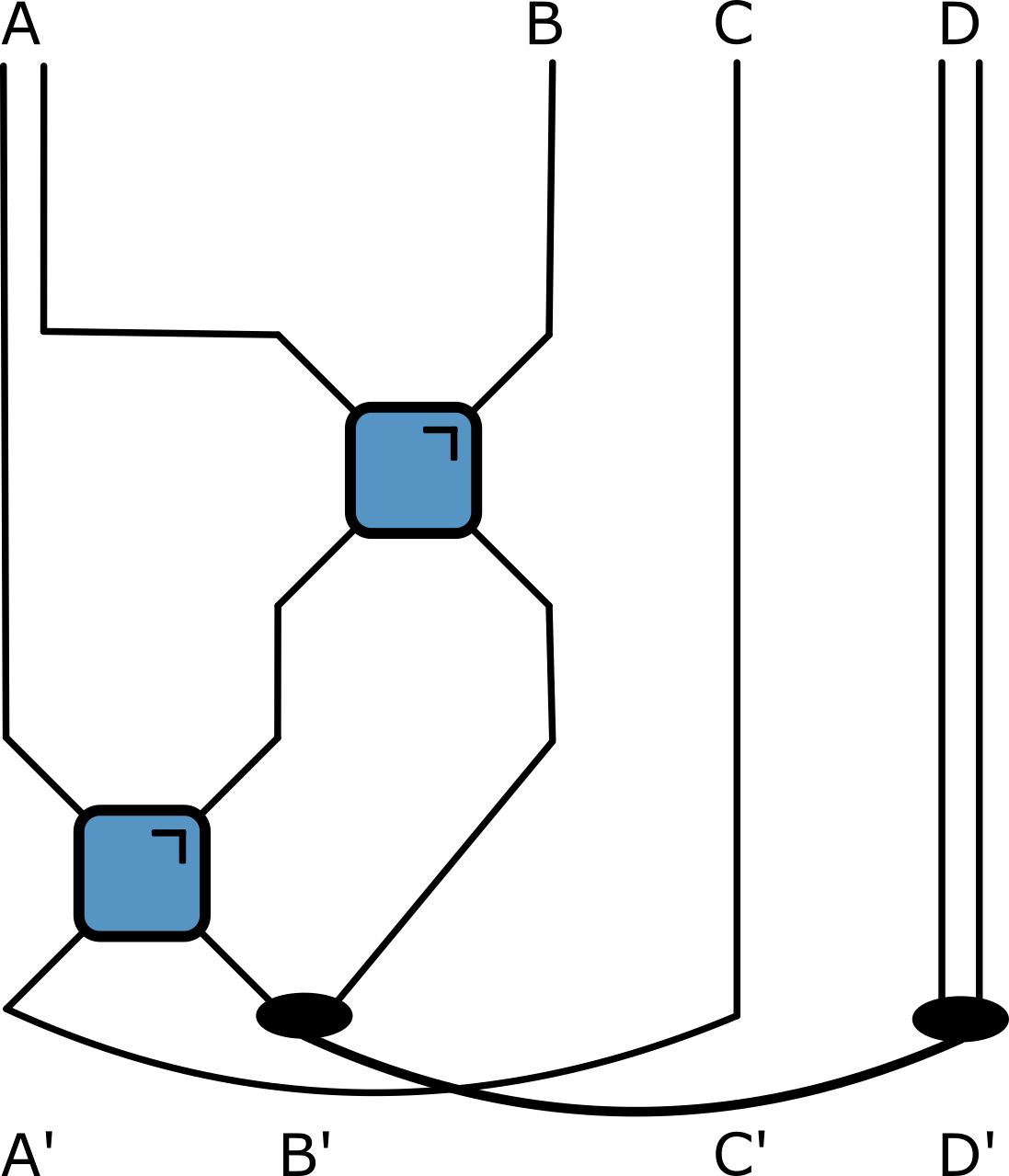}}} \label{eq:op_to_state_general}.
\end{align}
Eq.~\eqref{eq:higher_entanglement} expresses that the operator entanglement deriving from the particular Schmidt decomposition introduced above is given by computing the purity $E_k(\mathfrak{U})=1-\operatorname{Tr}[\rho_{AC}^2]$ with respect to the subset $AC$.

Consider what happens to a separable gate in this setup. Take e.g., $U$ to be the identity, for which direct computation returns $B_k=q^2$. The operator-to-state mapping can be represented as
\begin{align}
\ket{\psi_1}=\vcenter{\hbox{\includegraphics[width=0.2\columnwidth]{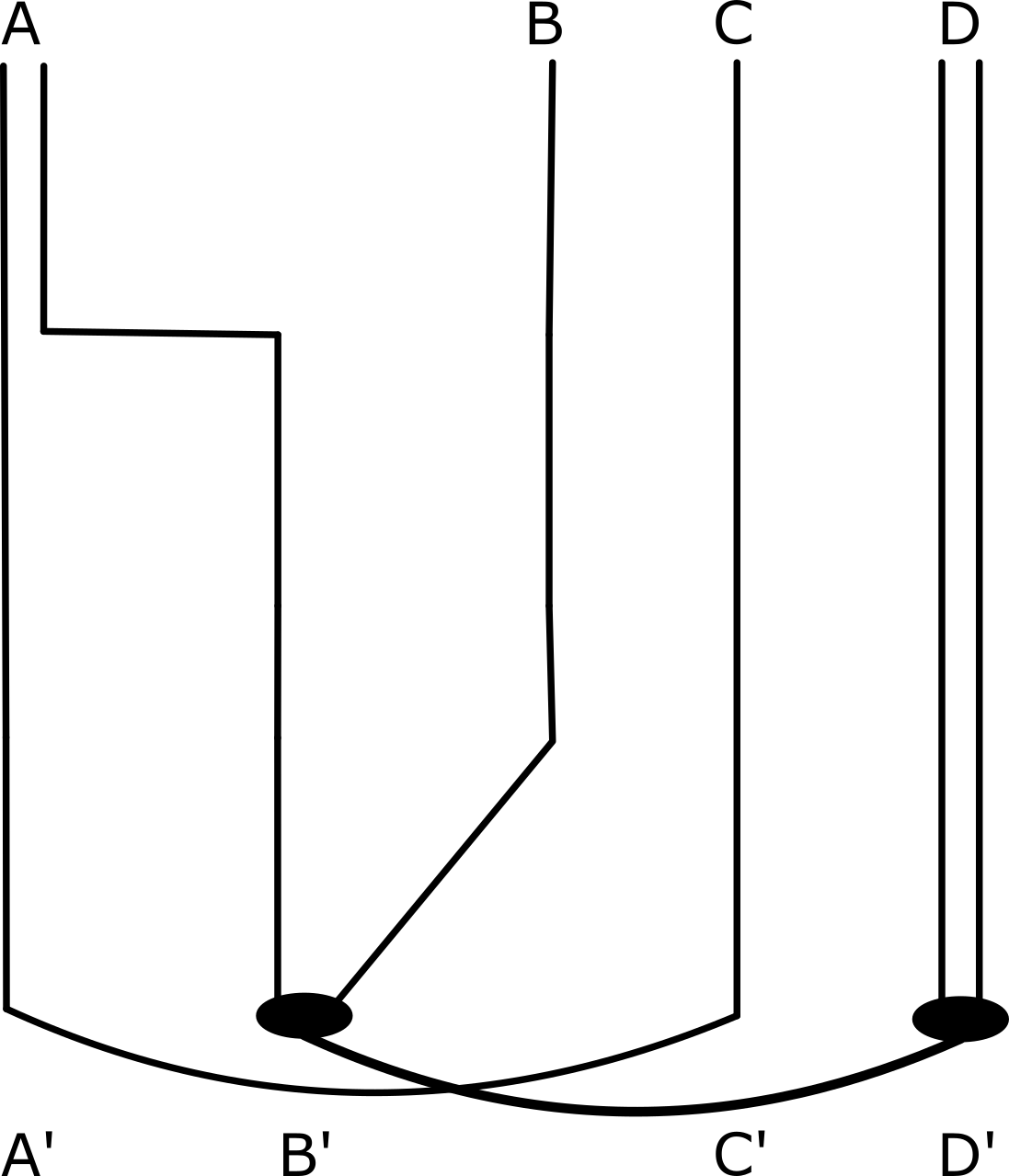}}} \label{eq:op_to_state_id}
\end{align}
The first site of $A$ and $C$ are maximally entangled, while the remaining sites of $A$ are independent of $C$. After tracing out $B'D'$ it holds that
\begin{equation}
    \rho_{AC} = \frac{1}{q^k}\sum_{n,m_1,...,m_{k-1}}\ket{nm_1\dots m_{k-1}n}\bra{nm_1\dots m_{k-1}n} = \ket{\phi}\bra{\phi}_{A_1 C} \otimes \frac{\mathbbm{1}}{q^{k-1}}. \label{eq:dm}
\end{equation}
This is exactly the product of a Bell state on sites $A_1$ and $C$ with a maximally mixed state on the remaining sites. It follows directly that $\operatorname{Tr}\rho_{AC}^2 = 1/q^{k-1}$, which gives $B_k=q^2$and reproduces the earlier result. It is a consequence of the gate not passing any entanglement between $A$ and $B$, thus $A$ and $C$ remain maximally entangled. In the general case, $U$ passes entanglement between $A$ and $B$ and thus, by monogamy of entanglement, the entanglement of $AC$ is diminished. Hence, $q^2$ indeed constitutes an uppper bound for $B_k$.

\subsection{Asymptotic Behavior for large $k$}

In this section we show that for gates that are not dual unitary $B_k\rightarrow q^2$ for $k\rightarrow\infty$. In the generic case, the convergence is determined by $B_1$.

First, by unfolding the graphical representation of $B_k$ it can be expressed as the contraction of a transfer matrix $T$
\begin{align}
B_k=\frac{1}{q}\,\vcenter{\hbox{\includegraphics[width=0.19\columnwidth,angle=-90]{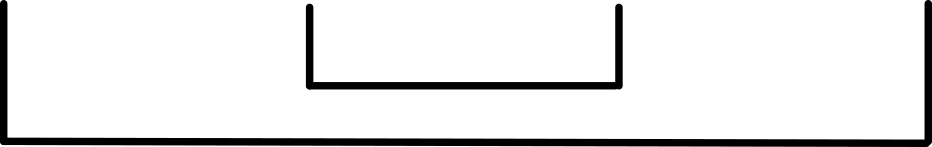}}}\left(\frac{1}{q}\vcenter{\hbox{\includegraphics[width=0.27\columnwidth,angle=-90]{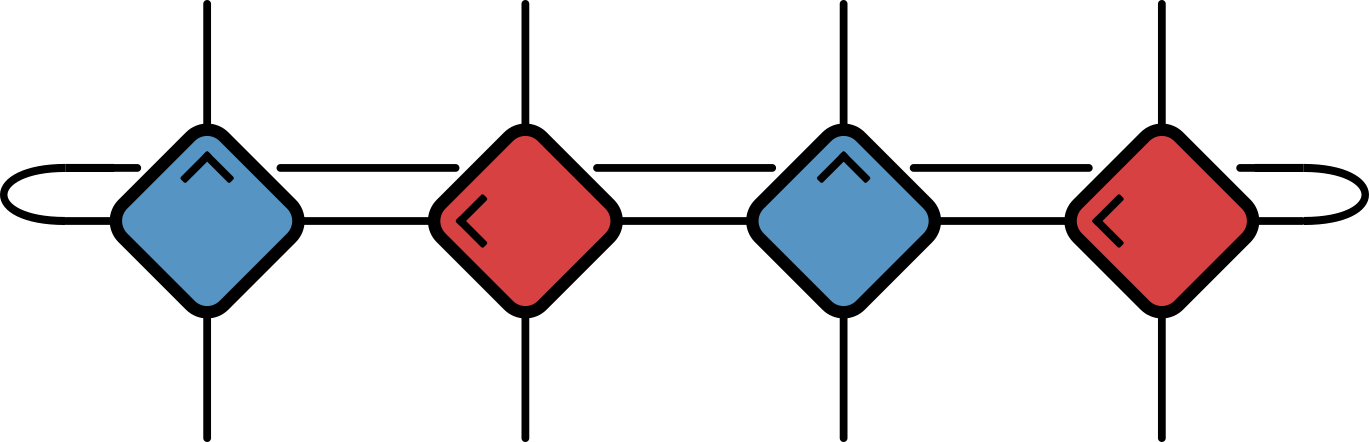}}}\right)^k\vcenter{\hbox{\includegraphics[width=0.19\columnwidth,angle=90]{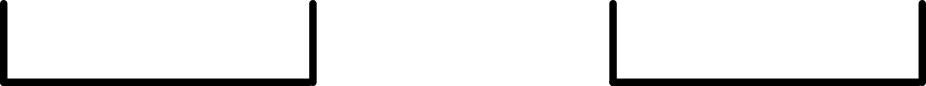}}} \label{eq:bk_tm}
\end{align}
Acting from the right with $T$ and interpreting the contraction on the left as a density operator, $T$ is a completely positive trace preserving map. Graphically 
\begin{align}
    T[\rho] = \frac{1}{q}\, \vcenter{\hbox{\includegraphics[width=0.13\columnwidth]{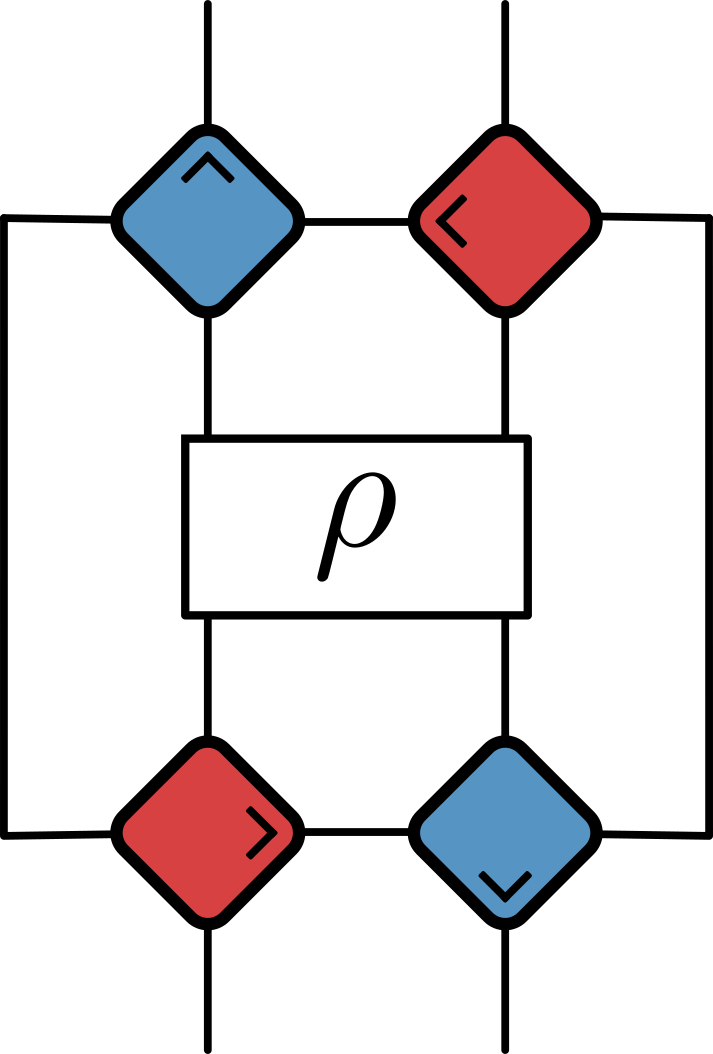}}}
\end{align}
This map allows for a direct operator sum representation as
\begin{align}
    T[\rho]_{aa',dd'} = \sum_{k,k'}E_{kk'}^{aa',bb'}\rho_{bb',cc'}E_{kk'}^{\dagger\,cc',dd'} \qquad \textrm{with} \qquad E_{kk'}^{aa',bb'} = \frac{1}{\sqrt{q}}U_{ka,bf}U^\dagger_{b'f,k'a'}, \label{eq:operatorsum}
\end{align}
in which $\sum_{k,k'}E_{kk'}^{\dagger\,cc',aa'}E_{kk'}^{aa',bb'}=\delta_{bc}\delta_{b'c'}$.
The contracting property of $T$ also immediately implies that $B_k\geq B_{k-1}$ and hence $z_k\geq0$.

Generically, this transfer matrix has one leading left- and right-eigenvector, equal to the transpose of the right and left boundary conditions respectively, and for $k\rightarrow\infty$ we can again replace the repeated application of the transfer matrix with a projection operator, now returning
\begin{align}
B_k\rightarrow\frac{1}{q^2}\,\vcenter{\hbox{\includegraphics[width=0.1\columnwidth]{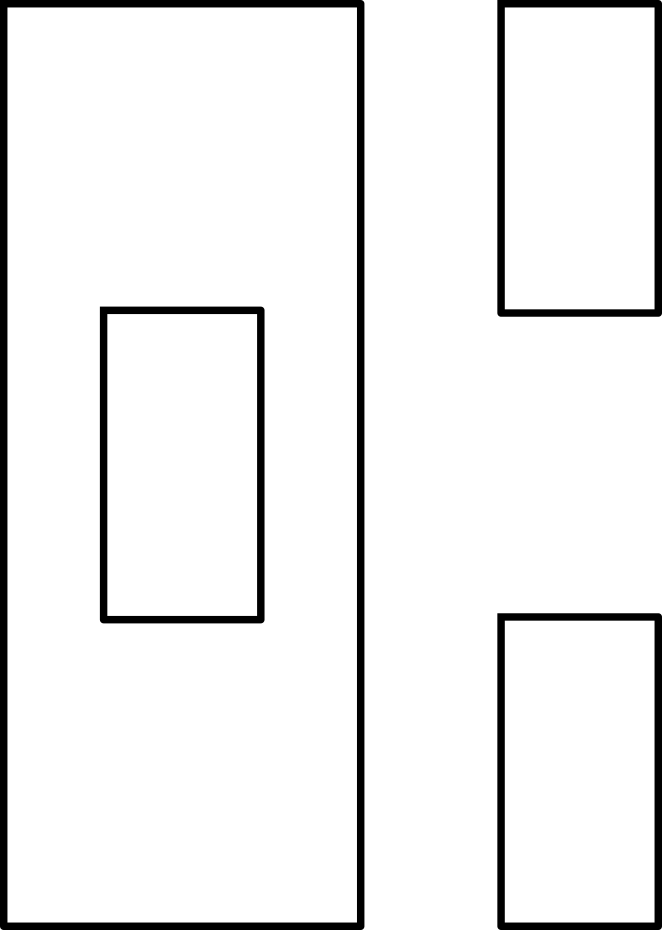}}}\,\,=q^2 \label{eq:bk_notdu}.
\end{align}
However, for dual unitary gates the transfer matrix has two degenerate leading eigenvectors, both left and right, such that $B_k=1$ for all $k$ since the boundary conditions are exact eigenstates, 
\begin{align}
\vcenter{\hbox{\includegraphics[width=0.14\columnwidth,angle=90]{figs/phi_R.png}}}\quad\mathrm{and}\quad\vcenter{\hbox{\includegraphics[width=0.14\columnwidth,angle=90]{figs/phi_L.png}}}\,\qquad \implies \qquad  B_k=\frac{1}{q}\,\,\vcenter{\hbox{\includegraphics[width=0.06\columnwidth]{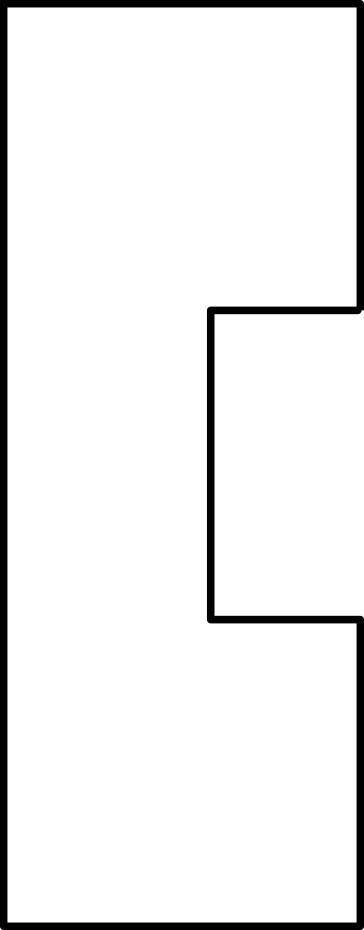}}}\,\,=1. \label{eq:bk_du}
\end{align}
This result can also be understood by noting that diagonal composition of dual-unitary gates preserves dual-unitarity \cite{Borsi2022}, such that we can immediately extend the argument that $B_1=1$ in the dual-unitary case to arbitrary values of $k$.

Close to dual unitarity we can use degenerate perturbation theory in this two-dimensional subspace to find
\begin{equation}
    B_k \approx q^2- (q^2-1)\left(1-\frac{B_1-1}{q^2-1}\right)^k. \label{eq:lower_bound}
\end{equation}
This approximate value is in fact a lower bound on $B_k$. In the representation of Eq.~\eqref{eq:operatorsum} the contraction from the left is equal to $T$ acting on the identity matrix. Since $T$ is a quantum channel, $T(\mathbbm{1})$ is a  positive operator which has a spectral decomposition $T(\mathbbm{1})=\sum_\alpha p_\alpha\ket{\psi_\alpha}\bra{\psi_\alpha}$ with $p_\alpha\geq0$ and the eigenvalues are identical to the eigenvalues of $T$. Consequently, we can write $B_k\sim \bra{\phi}T(\mathbbm{1})\ket{\phi}=\sum_\alpha p_\alpha\lvert\braket{\phi}{\psi_\alpha}\rvert^2$, with $\ket{\phi}$ a Bell state. Thus, the contribution of every single eigenvector in the spectral decomposition of $T$ to $B_k$ is nonnegative, and it follows that Eq.~\eqref{eq:lower_bound} constitutes a lower bound.  
Combining this lower bound with the previously established upper bound in turn implies an upper bound for $z_k$ via
\begin{equation}
    z_k = \frac{B_k-B_{k-1}}{q^2-1} \leq \frac{q^2-\left(q^2- (q^2-1)\left(1-z_1\right)^{k-1}\right)}{q^2-1} = (1-z_1)^{k-1}.
\end{equation}

\section{Path Integral Formula for the OTOC}

In this section the diagram rules for the computation of the OTOC are derived. These rules are then used to obtain analytic expressions for the OTOC after identifying the dominating contributions to the path integral. We discuss the one- and two-step approximations that result from restricting to paths containing steps of at most length 1 (2) and analyze their asymptotic behavior. We demonstrate that the one-step paths dominate the calculation of the OTOC in the limit of a large local Hilbert space.

\subsection{Derivation of the Diagram Rules}

We write the transfer matrix in the orthogonal basis of the MCS introduced in the first section as 
\begin{align}
    T_n &= D + \sum_{j=1}^n \rket{u_j}\rbra{j}, \qquad D =\operatorname{diag}\left(1,1-z_1,\dots,1-z_1 \right), \label{eq:decomp}
\end{align}
For convenience we introduce $M:=T_n-D = \sum_{j=1}^n \rket{u_j}\rbra{j}$. To compute the OTOC we need powers of $T_n$. As the matrix $M$ is linear in the scattering amplitudes $z_k$, we can expand $T_n^m$ in $M$ as (notice that we do not expand $D$ in $z_1$ -- we  comment on this in the next section) 
\begin{equation}
    T_n^m = D^m + \sum_{k=0}^{m-1} D^{k}M D^{m-1-k} + \sum_{\substack{k_0,k_1,k_2\geq0\\k_0+k_1+k_2=m-2}} D^{k_0} MD^{k_1} M D^{k_2} + \dots.
\end{equation}
The zeroth order term is given by
\begin{equation}
    C^{+(0)}(n,m) = -\frac{(1-z_1)^m}{q^2-1} \delta_{n,1}.
\end{equation}
For $z_1>0$ this expression differs from the dual-unitary result because the matrix $D$ already takes into account certain deviations from dual unitarity due to the inclusion of the $z_1$ terms in $D$. These terms can be interpreted as scattering processes that leave some weight of the right edge of the operator string on the edge of the light cone. Physically, they lead to damping: the right edge of the operator string becomes trivial after a time $\sim(\log(1-z_1))^{-1}$. We could have equally well absorbed these terms into the propagator by setting $(j\vert u_j)=-z_1$ (which explains the interpretation as scattering processes), at the expense of having paths with negative weights.  

The $\nu$-th order contribution can be generally written as
\begin{equation}
    T_n^{m\,(\nu)} = \sum_{\substack{k_0,\dots,k_\nu\geq0\\k_0+\dots+k_\nu=m-\nu}} D^{k_0} MD^{k_1} \dots M D^{k_\nu}.
\end{equation}
From the explicit form of $D$ if follows that $MD=(1-z_1) M$, which allows to write
\begin{equation}
    T_n^{m\,(\nu)} = \sum_{\substack{k_0,\dots,k_\nu\geq0\\k_0+\dots+k_\nu=m-\nu}} (1-z_1)^{m-\nu-k_0} D^{k_0} M^\nu = \sum_{k_0=0}^{m-\nu} \left(\sum_{\substack{k_1,\dots,k_\nu\geq0\\k_1+\dots+k_\nu=m-\nu-k_0}}\right) (1-z_1)^{m-\nu-k_0} D^{k_0} M^\nu.
\end{equation}
From this expression we immediately see that the effect of light-cone-edge scattering can be absorbed entirely into a renormalization of the first vertex $\rvert u_{j_1})$ that appears in each path, or alternatively speaking into the overlap with the left boundary $(L_n\vert u_{j_1})$. Notice that this overlap now depends on the coordinates $n,m$ and the order in perturbation theory $\nu$. The combinatorics of the edge scattering is contained in the generating function
\begin{equation}
    P_{m,\nu}(x) := \sum_{k_0=0}^{m-\nu} \left(\sum_{\substack{k_1,\dots,k_\nu\geq0\\k_1+\dots+k_\nu=m-\nu-k_0}}\right) x^{k_0}.
\end{equation}
The coefficients of the polynomial are given by the number of solutions to the equation $k_1+\dots+k_\nu=m-\nu-k_0$ where the $k_i$ are nonnegative integers and $k_0$ is held fixed. By ``stars and bars" it follows that
\begin{equation}
    P_{m,\nu}(x) = \sum_{k=0}^{m-\nu} \binom{m-k-1}{\nu-1} x^k.
\end{equation}
For $\nu\geq2$ this polynomial can be brought into the form of a hypergeometric function by defining $\beta_k = \binom{m-k-1}{\nu-1}/\binom{m-1}{\nu-1}$. Then
\begin{equation}
    \frac{\beta_{k+1}}{\beta_k} = \frac{(1+k)(\nu-m+k)}{(1+k)(1-m+k)},
\end{equation}
and consequently
\begin{equation}
    P_{m,\nu}(x) = \binom{m-1}{\nu-1}\hypf(1,-m+\nu;-m+1;x).
\end{equation}
The last piece required to compute  the renormalized left overlaps is the action of $D$ on the propagators, $D\vert u_j)=(u_{j0},(1-z_1) u_{j1},\dots,(1-z_1) u_{jn})$. Hence,
\begin{subequations}
\begin{align}
    (\Tilde{L}_{n,m,\nu}\vert u_{j_1=1}) &= \frac{(1-z_1)^{m-\nu}\sqrt{q^2-1}}{q^{n/2}}P_{m,\nu}((1-z_1)^{-1})z_1, \\
    (\Tilde{L}_{n,m,\nu}\vert u_{j_1\geq 1}) &= \frac{(1-z_1)^{m-\nu}\sqrt{q^2-1}}{q^{\frac{n}{2}+j_1-1}}\left(P_{m,\nu}((1-z_1)^{-1})z_j - P_{m,\nu}(1)\left(q^2z_{j-1}-z_j\right)\right).
\end{align}
\end{subequations}

The OTOC is given by computing the overlap
\begin{equation}
    C^\pm(n,m) = \sum_{\nu=0}^m \rbra{L_n}T_n^{m\,(\nu)}\rket{R_n^\pm} = \sum_{\nu=0}^m(1-z_1)^{m-\nu}\rbra{L_n}P_{m,\nu}\left((1-z_1)^{-1}D\right)M^\nu\rket{R_n^\pm}.
\end{equation}
By defining the renormalized left overlap $\rbra{\Tilde{L}_{n,m,\nu}}:=\rbra{L_n}P_{m,\nu}\left((1-z_1)^{-1}D\right)$ we can write this as a sum over weights of the form
\begin{subequations}
\begin{align}
    C^\pm(n,m) &= \sum_{\nu=0}^m(1-z_1)^{m-\nu} \sum_{j_1,\dots,j_\nu}W_{n,m,\nu}^\pm(j_1,\dots,j_\nu), \\
    W_{n,m,\nu}^\pm(j_1,\dots,j_\nu) &= (\Tilde{L}_{n,m,\nu}^Tu_{j_1})\left(\prod_{i=1}^{\nu-1}(j_i\vert u_{j_{i+1}})\right) (j_{\nu}\vert R^{\pm}_n).
\end{align}
\end{subequations}
The indices $\{j_1,\dots,j_\nu\}$ can be interpreted as nodes of a path and the quantities $(\ell\vert u_k)$ play the role of propagators determining the amplitude of jumping $k-\ell$ steps inside the light cone. The amplitudes for negative step sizes vanish making sure only causal paths contribute to the OTOC. This allows us to formulate the following diagram rules
\begin{itemize}
    \item For each $1\leq\nu\leq m$ consider all paths with $\nu$ nodes, $\{j_1,\dots,j_\nu\}$, such that $j_i<j_{i+1}$ (causality) and $j_\nu\leq n$.
    \item For a given path, associate each step $i\rightarrow i+1$ with the propagator $(j_i\vert u_{j_{i+1}})$, the starting point with $(\Tilde{L}_{n,m,\nu}\vert u_{j_1})$ and the endpoint with $(j_{\nu}\vert R^{\pm}_n)$.
    \item Sum over all paths at a given order, and then over all orders, weighted by $(1-z_1)^{m-\nu}$.
\end{itemize}

\subsection{Alternative Diagram Rules}

In the above approach, we have absorbed the amplitude for edge scattering processes into the diagonal matrix $D$. As a consequence, the effect of these processes is to renormalize the left boundary condition of the path integral. It is however also possible to keep these processes in the path integral, at the expense of negative weights and more involved combinatorics. In that case we have
\begin{subequations}
\begin{align}
    C^\pm(n,m) &= \sum_{\nu=0}^m\binom{m}{\nu} \sum_{j_1,\dots,j_\nu}W_n^{'\pm}(j_1,\dots,j_\nu), \\
    W_n^{'\pm}(j_1,\dots,j_\nu) &= (L_n\vert u'_{j_1})\left(\prod_{i=1}^{\nu-1}(j_i\vert u'_{j_{i+1}})\right) (j_{\nu}\vert R^{\pm}_n),
\end{align}
\end{subequations}
where $(j\vert u'_{j})=-z_1$ and $(j\vert u'_{k})=(j\vert u_{k})$ for $j\neq k$. The diagram rules now read
\begin{itemize}
    \item For each $1\leq\nu\leq m$ consider all paths with $\nu$ nodes, $\{j_1,\dots,j_\nu\}$, such that $j_i\leq j_{i+1}$ and $j_\nu\leq n$.
    \item For a given path, associate each step $i\rightarrow i+1$ with the propagator $(j_i\vert u'_{j_{i+1}})$, the starting point with $(L_{n}\vert u'_{j_1})$ and the endpoint with $(j_{\nu}\vert R^{\pm}_n)$.
    \item Sum over all paths at a given order, and then over all orders, weighted by $\binom{m}{\nu}$.
\end{itemize}

\subsection{one-step Approximation}

Using the diagram rules, the OTOC in the one-step approximation can be found. Only two distinct paths contribute. The first starts at $j_1=1$ followed by $n-1$ steps upward ($\nu=n$), and the second starts at $j_1=2$ followed by $n-2$ steps upward ($\nu=n-1$). This yields
\begin{equation}
    C_{(1)}^+(n,m) = z_1^n (1-z_1)^{m-n} P_{m,n}(1-z_1) - \frac{1}{q^2-1}z_1^{n-1} (1-z_1)^{m-n+1} P_{m,n-1}(1). \label{eq:c1}
\end{equation}
We define the front profile in the large $q$ limit
\begin{equation}
    F_{z_1}(n,m) := z_1^n(1-z_1)^{m-n} P_{m,n}(1-z_1).
\end{equation}
We find that the second term in Eq.~\eqref{eq:c1} vanishes asymptotically, while the first, given by $F_{z_1}$, remains finite.

The above considerations have to be modified exactly on the light cone, $n=1$, where paths cannot start at $2$ and the zeroth order term $\rbra{L_n}D^m\rket{R_n^+}$ also contributes. Explicit calculation yields
\begin{equation}
    C^+(n=1,m) = 1 - \frac{q^2}{q^2-1}(1-z_1)^m.
\end{equation}

\subsubsection{Asymptotic Analysis of the Front Profile}

While we have an exact expression for the OTOC in the MCS in the large $q$ limit, the function is inefficient to compute numerically for large values of its argument. To better understand the behavior of the OTOC, we investigate the properties of $F_{z_1}(n,m)$ for large $n,m$.

First, we can rewrite $F_{z_1}$ in terms of the incomplete beta function $B_{z_1}(p,q)$
\begin{equation}
    F_{z_1}(n,m)=n \binom{m}{n}B_{z_1}(n,m-n+1).
\end{equation}
The following asymptotic analysis is facilitated by using the integral representation of the incomplete beta function,
\begin{equation}
    B_{z_1}(p,q) = \int_0^{z_1} \mathrm{d}t\, t^{p-1}(1-t)^{q-1}\,.
\end{equation}

To determine the butterfly velocity we investigate the OTOC on rays of constant velocity $x=vt$ as $t\rightarrow\infty$. Recall that $n,m$ are related to the spacetime coordinates as
\begin{equation}
    n = \frac{t-x+2}{2} = \frac{1-v}{2}t+1, \quad m = \frac{t+x}{2} = \frac{1+v}{2}t.
\end{equation}
Hence, in the relevant limit, $\kappa^{-1}:=m/(n-1)=(1+v)/(1-v)$ is constant. We write
\begin{align}\label{eq:sm:B_int}
    B_{z_1}(n,m-n+1) &= \int_0^{z_1} \mathrm{d}s\, s^{n-1}(one-s)^{m-n} = \int_0^{z_1} \mathrm{d}s \frac{1}{one-s} \exp\left( m\left( \log(one-s) + \frac{n-1}{m}\log\left(\frac{s}{one-s}\right) \right)\right) \\
    &=\int_0^{z_1} \mathrm{d}s \frac{1}{one-s} \exp\left( m  g(s) \right)\,,
\end{align}
in which we have defined
\begin{align}
    g(s):=\log(one-s) + \frac{n-1}{m}\log\left(\frac{s}{one-s}\right)\,.
\end{align}
Eq.~\eqref{eq:sm:B_int} has the form of an Laplace integral, which can be asymptotically analyzed using a saddle-point approximation. We find that the saddle point is situated at $s_0=\kappa=\frac{1-v}{1+v}$. For fixed perturbation strength $z_1$ the saddle point is located either inside, outside, or on the border of the integration region depending on the value of the velocity. The asymptotic behavior on rays $x=vt$ thus drastically changes its character as a critical velocity $v_{B,1}=\frac{1-z_1}{1+z_1}$ is crossed. This velocity corresponds exactly to the butterfly velocity, as will be demonstrated below.

Consider first the case $v<v_{B,1}$ in which the saddle point lies outside the integration domain. Physically we expect the OTOC to decay exponentially to zero as $t\rightarrow\infty$, signifying scrambling. The large $n$ approximation to the beta function reads
\begin{equation}
    B_{z_1}(n,m-n+1) \approx \frac{e^{mg(z_1)}}{m(1-z_1)g'(z_1)} = \frac{(1-z_1)^{m-n+1}z_1^n}{m\left(\frac{n-1}{m}-z_1\right)}.
\end{equation}
We further use Stirling's approximation for the binomial
\begin{equation}
    \binom{m}{n} \approx \sqrt{\frac{m}{2\pi(m-n)n}} \frac{m^m}{(m-n)^{m-n}n^n} \approx \frac{2v}{1-v}\sqrt{\frac{1+v}{2\pi(1-v) vt}} \zeta^t,
\end{equation}
where 
\begin{equation}
    \log\zeta(v) := \frac{1+v}{2}\log\frac{1+v}{2} - \frac{1-v}{2}\log\frac{1-v}{2} -v\log v. \label{eq:zeta_exponential}
\end{equation}
Putting these results together we find 
\begin{equation}
    C_{(1)}^+( x=vt,t) \approx \frac{2v}{1+v} \frac{z_1}{\frac{1-v}{1+v}-z_1}\sqrt{\frac{1+v}{2\pi(1-v) vt}} \left[\zeta (1-z_1)^v z_1^{\frac{1-v}{2}} \right]^t \sim \frac{\gamma(v,z_1)^t}{\sqrt{t}},
\end{equation}
with $\gamma(v,z_1)$ given by
\begin{equation}
    \gamma(v,z_1) = \zeta(v) (1-z_1)^v z_1^{\frac{1-v}{2}}\,.
\end{equation}
For $v<(1-z_1)/(1+z_1)$ we have that $\gamma$ is smaller than $1$, such that the OTOC decays exponentially inside the light cone, as expected. We define a scrambling time as $t_\ast:=-(\log\gamma(v,z_1))^{-1}$. Moreover, we have $\gamma\left(v=\frac{1-z_1}{1+z_1},z_1\right)=1$. 

Outside the light cone, for $v>v_B$, the saddle point lies inside the integral and hence the large $n$ analysis yields
\begin{equation}
    B_{z_1}(n,m-n+1) \approx \frac{e^{mg(\kappa)}}{1-\kappa}\sqrt{\frac{2\pi}{m\abs{g''(\kappa)}}} = \sqrt{\frac{\pi \kappa(1-\kappa)}{2m}}(1-\kappa)^{m-n}\kappa^{n-1}.
\end{equation}
As the condition $\gamma(v,\kappa)=1$ is satisfied on the saddle point, the OTOC approaches a constant
\begin{equation}
    C_{(1)}^+( x=vt,t) \approx 1.
\end{equation}
This indicates that operators outside the light cone commute.

These two results together already show that $v_{B,1}$ indeed equals the butterfly velocity. In the following the form of the OTOC close to the front, $x-v_{B,1}t=\mathcal{O}(\sqrt{t})$, is determined. We write $x=v_{B,1} t+\delta$ where we assume $\delta\sim\mathcal{O}(\sqrt{t})$. We begin with evaluating the incomplete $\beta$-function in this limit. The integral is no longer a Laplace integral, but it possesses two large parameters. Following Fulks' method from Ref.~\cite{Fulks1951}, we find that (recall that the saddle point of the original problem lies on the boundary of the interval) \footnote{Formally Fulks' method is only applicable for $\delta>0$, but numerical comparisons show that this approximation remains accurate for $\delta<0$.} 
\begin{equation}
    B_{z_1} \approx A(t)\left(\frac{2v_{B,1}}{\sqrt{1-v_{B,1}^2}}\right)^\delta e^{\frac{\delta^2}{2v_{B,1}(1-v_{B,1}^2)t}} \left( 1 + \operatorname{erf}\left(\frac{\delta}{\sqrt{2v_{B,1}(1-v_{B,1}^2)t}}\right)\right) + \mathcal{O}(t^{-1/2}),
\end{equation}
where we have subsumed the known prefactors only depending on $t$ in the function $A(t)$, and have introduced the error function $\operatorname{erf}(x)=\frac{2}{\sqrt{\pi}}\int_0^x\mathrm{d}s\,e^{-s^2}$. Let us also analyze the remaining terms in $F$. The prefactor contains only terms that scale as $\delta/t\sim1/t^{1/2}$ beyond leading order. The binomial calls for a more thorough analysis. Formally, Fulks' method has to also be applied, but in this case this approach can be avoided by carefully expanding Stirling's approximation in $\delta$. Corrections to Stirling's approximation are not relevant, since they are suppressed by at least $1/t$. Dropping terms of order $1/t$, we write
\begin{align}
    \binom{m}{n} \approx\,& 2\sqrt{\frac{((1+v_{B,1})t+\delta)(v_{B,1}t+\delta)}{2\pi((1-v_{B,1})t-\delta)^3}} \left[\frac{\left(\frac{(1+v_{B,1})t+\delta}{2}\right)^{(1+v_{B,1})/2}}{\left(\frac{(1-v_{B,1})t-\delta}{2}\right)^{(1-v_{B,1})/2}\left(v_{B,1}t+\delta\right)^{v_{B,1}}}\right]^t \nonumber\\
    & \times \left[\frac{\left(\frac{(1+v_{B,1})t+\delta}{2}\right)^{1/2}\left(\frac{(1-v_{B,1})t-\delta}{2}\right)^{1/2}}{v_{B,1}t+\delta}\right]^\delta.
\end{align}

The different contributions to the exponentials can now be considered. At leading order we can pull out $\zeta^t$, with $\zeta$ defined in Eq.~\eqref{eq:zeta_exponential}. The remaining term reads
\begin{align}
    &\exp\left[t\left(\frac{1+v_{B,1}}{2}\log\left(1+\frac{\delta}{(1+v_{B,1})t}\right) - \frac{1-v_{B,1}}{2}\log\left(1-\frac{\delta}{(1-v_{B,1})t}\right) - v_{B,1}\log\left(1+\frac{\delta}{v_{B,1}t}\right)\right)\right] \nonumber \\
    &\qquad \approx \exp\left(\frac{\delta^2}{2v_{B,1}(1-v_{B,1}^2)t} + \mathcal{O}(1/t^{1/2})\right).
\end{align}
A similar analysis of the exponential in $\delta$ yields
\begin{equation}
    \left[\frac{\left(\frac{(1+v_{B,1})t+\delta}{2}\right)^{1/2}\left(\frac{(1-v_{B,1})t-\delta}{2}\right)^{1/2}}{v_{B,1}t+\delta}\right]^\delta \approx \left[\frac{\sqrt{1-v_{B,1}^2}}{2v_{B,1}}\right]^\delta \exp\left(-\frac{\delta^2}{v_{B,1}(1-v_{B,1}^2)t}\right).
\end{equation}
These are now all terms that are of order one in the limit $t\rightarrow\infty,\,\delta\sim\sqrt{t}$. Hence we conclude
\begin{equation}
    C_{(1)}^+( x=v_{B,1}t+\delta,t) \approx \frac{1}{2}\left(1 + \operatorname{erf}\left(\frac{\delta}{\sqrt{2v_{B,1}(1-v_{B,1}^2)t}}\right)\right) + \mathcal{O}(1/t^{1/2}). \label{eq:front_asymptotic}
\end{equation}
This is indeed the shape of a diffusively broadening front, $\frac{1}{2}(1+\operatorname{erf}(\frac{\delta}{\sqrt{2D_1t}}))$, with diffusion constant
\begin{equation}
    D_1 = (1-v_{B,1}^2)v_{B,1}\,.
\end{equation}

\subsection{two-step Approximation}

Using the above rules, the OTOC in the two-step approximation can be constructed. There are three possible starting points. Starting at $n=1$ we denote the number of two-steps by $h_2$. There are $\binom{n-1-h_2}{h_2}$ choices to distribute these steps. A given path with $h_2$ two-steps has a total of $\nu=n-h_2$ nodes. We find
\begin{align}
    C_{(2),n_0=1}^+ = &\frac{z_1}{q^{n-1}}  \sum_{h_2=0}^{\lfloor\frac{n-1}{2}\rfloor} \binom{n-1-h_2}{h_2}\left(\frac{q^2z_2}{q^2}\right)^{h_2} \times\dots\nonumber\\
    &\times\left(\frac{q^2z_1-z_2}{q}\right)^{n-1-2h_2} \left(1-z_1\right)^{m-n +h_2} P_{m,n-h_2}\left(1-z_1\right).
\end{align}
Introducing $\Tilde{z}_1=z_1-z_2/q^2$ we can rewrite this in terms of the first-order front profile $F_{z_1}$
\begin{equation}
    C_{(2),n_0=1}^+ = \left(\frac{\Tilde{z}_1}{z_1}\right)^{n-1} \sum_{h_2=0}^{\lfloor\frac{n-1}{2}\rfloor} \binom{n-1-h_2}{h_2} \left(\frac{z_2z_1}{q^2\Tilde{z}_1^2}\right)^{h_2} F_{z_1}(n-h_2,m).
\end{equation}
This is a weighted sum over fronts shifted to $n\rightarrow n-h_2$. In coordinate space this means that the front is shifted in the $(t-x)$-direction. We can express this using only two microscopic parameters, $z_1$ and the combination $\xi=\frac{z_2}{q^2 z_1}$, since it holds that
\begin{equation}
    \frac{\Tilde{z}_1}{z_1} = 1-\frac{z_2}{q^2 z_1} = 1-\xi, \qquad \frac{z_2z_1}{q^2\Tilde{z}_1^2} = \frac{z_1}{\Tilde{z}_1}\frac{z_2}{q^2(z_1-q^{-2}z_2)} = \frac{\xi}{(1-\xi)^2}.
\end{equation}
Notice that $\xi$ vanishes in the large $q$ limit which implies that in this limit the two-step contribution becomes unimportant (assuming that all the $z_k$ remain finite as implied by the maximal $B_k\sim q^2$ scaling).

The remaining contributions can similarly be found as
\begin{subequations}
    \begin{align}
        C_{(2),n_0=2}^+ =& \frac{z_2}{q^2z_1} \left(\frac{\Tilde{z}_1}{z_1}\right)^{n-2} \sum_{h_2=0}^{\lfloor\frac{n-2}{2}\rfloor} \binom{n-2-h_2}{h_2} \left(\frac{z_2z_1}{q^2\Tilde{z}_1^2}\right)^{h_2} F_{z_1}(n-1-h_2,m) \nonumber\\
        &\qquad- \frac{\Tilde{{z_1}}^{n-1}}{q^2-1} \sum_{h_2=0}^{\lfloor\frac{n-2}{2}\rfloor} \binom{n-2-h_2}{h_2} \left(\frac{z_2}{q^2\Tilde{z}_1^2}\right)^{h_2} (1-z_1)^{m-n+1+h_2} P_{m,n-1-h_2}(1)\,, \\
        C_{(2),n_0=3}^+ =& - \frac{z_2 \Tilde{z}_1^{n-3}}{q^2} \sum_{h_2=0}^{\lfloor\frac{n-3}{2}\rfloor} \binom{n-3-h_2}{h_2} \left(\frac{z_2}{q^2\Tilde{z}_1^2}\right)^{h_2} (1-z_1)^{m-n+2+h_2} P_{m,n-2-h_2}(1)\,.
    \end{align}
\end{subequations}

In the following we present an asymptotic expansion of these results that can be used to extract the butterfly velocity and diffusion constant.  We start with $C_{(2),n_0=1}^+$, noting that $C_{(2),n_0=2}^+$ is analogous and $C_{(2),n_0=3}^+$ vanishes asymptotically. First, we introduce the continuum variable $h$ through $h_2=h(n-1)$. We can approximate $C_{(2),n_0=1}^+$ as an integral and insert the asymptotic expression for the binomial
\begin{equation}
    C_{(2),n_0=1}^+ \approx (n-1) \int_0^{1/2}\mathrm{d}h \frac{1}{\sqrt{2\pi(n-1)}}\sqrt{\frac{1-h}{h(1-2h)}}e^{(n-1)H(h)}F_{z_1}((1-h)(n-1)+1,m),
\end{equation}
where
\begin{equation}
    H(h) = (1-h)\log(1-h)-h\log(h)-(1-2h)\log(1-2h) + h\log\left(\frac{\xi}{(1-\xi)^2}\right) +\log(1-\xi).
\end{equation}
Finding the maximum of $H$ reduces to an algebraic equation solved by
\begin{equation}
    h_{\mathrm{max},\pm} = \frac{1}{2}\left(1\pm\frac{1}{\sqrt{1+\frac{4\xi}{(1-\xi)^2}}}\right).
\end{equation}
Only the $h_{\mathrm{max},-} $ solution lies inside the domain of integration. Crucially, it holds that $H(h_{\mathrm{max},-})=0$, such that the large $n$ analysis yields
\begin{subequations}
\begin{align}
    C_{(2),n_0=1}^+ &\approx (1-\xi) F_{z_1}\left( (1-h_{\mathrm{max}})(n-1)+1,m\right),\\
    C_{(2),n_0=2}^+ &\approx \xi F_{z_1}\left( (1-h_{\mathrm{max}})(n-2)+1,m\right).
\end{align}
\end{subequations}
The $C_{(2),n_0=2}^+$ contribution is dominated by the same saddle point, and $C_{(2),n_0=3}^+$ vanishes asymptotically. Overall, the front function remains the same, the higher-order contribution only serves to renormalize the arguments.

In order to extract $v_B$ and $D$, we consider $F_{z_1}$ in the regime where it can be approximated by an error function. The argument of the error function now reads
\begin{equation}
    \frac{(1-(1+v_{B,1})\frac{\hmax}{2})x-(v_{B,1}-(1+v_{B,1})\frac{\hmax}{2})t}{\sqrt{2D_1\left((1-\frac{\hmax}{2})t+\frac{\hmax}{2}x\right)}},
\end{equation}
and we read off
\begin{subequations}
    \begin{align}
        v_{B,2} &= \frac{v_{B,1}-(1+v_{B,1})\frac{\hmax}{2}}{1-(1+v_{B,1})\frac{\hmax}{2}} \approx v_{B,1} - \frac{1-v_{B,1}^2}{2}\hmax,\\
        D_2 &= D_1\frac{1-(1-v_{B,2})\frac{\hmax}{2}}{\left(1-(1+v_{B,1})\frac{\hmax}{2}\right)^2} \approx D_1 \left(1+\frac{1+3v_{B,1}}{2}\hmax\right).
    \end{align}
\end{subequations}

\subsection{Larger Steps}

We argue that close to the front the inclusion of $k$-steps with $k>1$ only renormalizes the parameters of the profile but does not change its functional form. Moreover, all contributions of higher steps vanish in the limit of large local Hilbert space when the scattering amplitudes $z_k$ are held constant.

Taking into account up to $k$-step processes we have asymptotically finite terms of the form
\begin{align}
    C_{(k)}^+(n,m) \sim \frac{1}{q^n}\sum_{\substack{h_2,\dots,h_k\\\sum jh_j\leq n-n_0}}& \mathcal{C}_n(h_2,\dots,h_k)\left(\frac{q^2z_k}{q^{k}}\right)^{h_k} \left(\frac{q^2z_{k-1}-z_k}{q^{k-1}}\right)^{h_{k-1}}\times\nonumber\\&\times\left(\frac{q^2z_1-z_2}{q}\right)^{n-2h_2-\dots-kh_k} F_{z_1}\left(n-\sum_{j=2}^k(j-1)h_j,m\right),
\end{align}
where $\mathcal{C}_n(h_2,\dots,h_k)$ is an unknown combinatorial factor. This can be interpreted as a polynomial in the variables $(q^2z_j-z_{j+1})/q^{j}$. It can be rewritten to reveal the dependence on $q$
\begin{align}
    &\left(\frac{q^2z_k}{q^{k}}\right)^{h_k} \left(\frac{q^2z_{k-1}-z_k}{q^{k-1}}\right)^{h_{k-1}}\times\dots\times\left(\frac{q^2z_1-z_2}{q}\right)^{n-2h_2-\dots-kh_k}  \nonumber\\
    \sim&\left(\frac{z_k}{q^{k-2}}\right)^{h_k}\left(\frac{\Tilde{z}_{k-1}}{q^{k-3}}\right)^{h_{k-1}}\times\dots\times\left(q\Tilde{z}_{1}\right)^{n-2h_2-\dots-kh_k}\nonumber\\
    \sim&\left(\frac{z_k}{q^{2(k-1)}\Tilde{z}_{1}^k}\right)^{h_k}\times\dots.
\end{align}
For $q^2\gg 1/\Tilde{z}_1\approx 1/z_1$ all these arguments become small and the 1st-order result is unaffected.

\end{widetext}

\end{document}